\documentclass[preprint]{aastex}

\usepackage{color}
\usepackage{gensymb}

\providecommand{\e}[1]{\ensuremath{\times 10^{#1}}}

\def\teff{\mbox{$T_{\rm eff}$}}
\def\logg{\mbox{log({\it g})}}
\def\vmicro{\mbox{$\xi_{\rm t}$}}
\def\kmsec{\mbox{km~s$^{\rm -1}$}}
\def\ciso{\mbox{$^{12}$C/$^{13}$C}}

\def\eg{\mbox{e.g.}}
\def\ew{\textit{EW}}
\providecommand{\e}[1]{\ensuremath{\times 10^{#1}}}

\shorttitle{ATMOSPHERES AND COMPOSITIONS OF M68 STARS}
\shortauthors{Schaeuble et al.}

\begin{document}
       
\title{A DETAILED STUDY OF GIANTS AND HORIZONTAL BRANCH STARS IN M68:
ATMOSPHERIC PARAMETERS AND CHEMICAL ABUNDANCES}

\author{M. Schaeuble\altaffilmark{1},
        G. Preston\altaffilmark{2}, 
        C. Sneden\altaffilmark{1},
        I. B. Thompson\altaffilmark{2},
        S. A. Shectman \altaffilmark{2},
        G. S. Burley \altaffilmark{2}}

\altaffiltext{1}{Department of Astronomy and McDonald Observatory,
                 The University of Texas, Austin, TX 78712;
                 mschaeu, chris@astro.as.utexas.edu}

\altaffiltext{2}{Carnegie Observatories,
                 813 Santa Barbara Street, Pasadena, CA 78712;
                 gwp, ian, shec, burley@obs.carnegiescience.edu}

\clearpage
\begin{abstract}
In this paper, we present a detailed high-resolution spectroscopic study 
of post main sequence stars in the Globular Cluster M68. Our sample, 
which covers a range of 4000 K in \teff, and 3.5 dex in \logg, is comprised
of members from the red giant, red horizontal, and blue horizontal branch, 
making this the first high-resolution globular cluster study covering such a large evolutionary
and parameter space. Initially, atmospheric parameters were determined using
photometric as well as spectroscopic methods, both of which resulted in 
unphysical and unexpected \teff, \logg, \vmicro, and [Fe/H] combinations.
We therefore developed a hybrid approach that addresses most
of these problems, and yields atmospheric parameters that agree
well with other measurements in the literature. Furthermore, our
derived stellar metallicities are consistent across all 
evolutionary stages, with
$\langle$[Fe/H]$\rangle$ = $-$2.42 ($\sigma$ = 0.14) from 25 stars.
Chemical abundances obtained using our methodology 
also agree with previous studies and bear all 
the hallmarks of globular clusters, such as a Na-O anti-correlation, constant 
Ca abundances, and mild $r$-process enrichment.

\end{abstract}

\clearpage
\section{INTRODUCTION\label{intro}}
Along with M15 and M92, M68 is one of the lowest metallicity 
Galactic globular clusters (GC), [Fe/H]~$\sim~$-2.3\footnote{
We adopt the standard spectroscopic notation
that for elements A and B,
[A/B] $\equiv$ log$_{\rm 10}$(N$_{\rm A}$/N$_{\rm B}$)$_{\star}$ $-$
log$_{\rm 10}$(N$_{\rm A}$/N$_{\rm B}$)$_{\odot}$.
We define
log~$\epsilon$ $\equiv$ log$_{\rm 10}$(N$_{\rm A}$/N$_{\rm H}$) + 12.0, and
equate metallicity with the stellar [Fe/H] value.} \cite{harris_catalog_1996}.
It has therefore been included in many large-sample light-element abundance 
studies (\eg, \citealt{bellazzini_na-o_2012}), but has thus far been subjected to very few 
detailed chemical composition investigations.
In an analysis of red horizontal branch (RHB) stars in the metal-poor
GC M15, \cite{preston_atmospheres_2006} found
a declining difference between surface gravities determined from photometry
and LTE spectrum analysis with increasing effective temperature 
in the range 5300~K \textless \ \teff \ \textless \ 6300~K, a temperature range
which embraces almost the entire RHB of that cluster. 
Contemporaneously, \citeauthor{lee_chemical_2005} (2005, hereafter Lee05) 
found larger differences in surface gravities among the cooler 
(\teff $\ \sim $\ 4200~K) red giant branch (RGB) stars of M68.  
In combination, these results suggest that non-LTE (NLTE) over-ionization 
of neutral metals produces systematic errors in abundance analyses of 
cool, metal-poor red giants in globular clusters, and that abundances 
derived from RHB stars may provide a more accurate abundance scale
for metal-poor stars of globular clusters and the Galactic halo. 

We decided to pursue this possibility by conducting an expanded 
investigation of post main sequence stars in M68, similar to that for M15 
reported by \citeauthor{sobeck_abundances_2011} (2011, hereafter Sob11).
From a purely observational point of view, M68 has 
many advantages over M15. 
First, the RR Lyrae stars in M68 are $\ \sim $0.2 mag. brighter
than those in M15 (\citeauthor{walker_bvi_1994} (1994, hereafter Wal94)).
Second, all of our observations were done at the Las Campanas Observatory 
(LCO), and since M68 has $\delta $\ = $-$27\degree, it transits the meridian 
near the zenith, whereas M15, with $\delta $\ = +12\degree, lies low in 
LCO's northern sky. See Table \ref{fun_params} for more fundamental parameters
of M68.

In addition to being more accessible observationally, M68 differs from M15 
in several other important respects. 
M68 has a relatively sparsely populated RHB compared to M15, which 
possesses an extended blue horizontal branch (EHB) 
(\citealt{durrell_color-magnitude_1993}). 
This is also reflected in the horizontal branch ratio, defined to be the 
ratio of horizontal branch stars to red giant stars, for both clusters. 
For M15, this ratio is 0.67, while for M68 it's only 0.17
(\citealt{zinn_globular_1986}, \citealt{harris_catalog_1996}).
Such differences in HB morphology have 
been studied extensively (e.g. \citealt{lee_horizontal-branch_1994}
and \citealt{caputo_helium_1980}) and attributed to variations
in cluster age and metallicity, as well as stellar helium
abundances and rotation (see Figs. 2-5 in
\citealt{lee_horizontal-branch_1994}). Since
M15 and M68 have very similar ages and metallicities, 
the differences in HB morphology could be attributed
to disparate He contents.

M68 and M15 also seem to exhibit systematically different abundance patterns. 
Lee05 found
$\langle\mathrm{[Si/Fe]\rangle_{M68}}\sim 0.6$,
while $\langle\mathrm{[Si/Fe]}\rangle_{\mathrm{M15}}\sim$ 0.2
(Sob11). Si abundances in `normal' Pop~II stars are
equivalent to those of M68 RGBs (Lee05, \citealt{cayrel_first_2004}).
Furthermore, M68 seems underabundant in Ti (Lee05), 
whereas overabundances of neutron capture elements, which vary from star
to star, are found in the RHB and RGB stars of M15.
Finally, M15 contains dusty red giants (\citet{boyer_stellar_2006}), 
surprising in view 
of the low metallicity, and interesting as unusually large 
mass loss during post main sequence evolution has been advanced as 
an explanation for the EHB (\citealt{dcruz_origin_1996}).

\begin{figure}
\epsscale{0.8}
\plotone{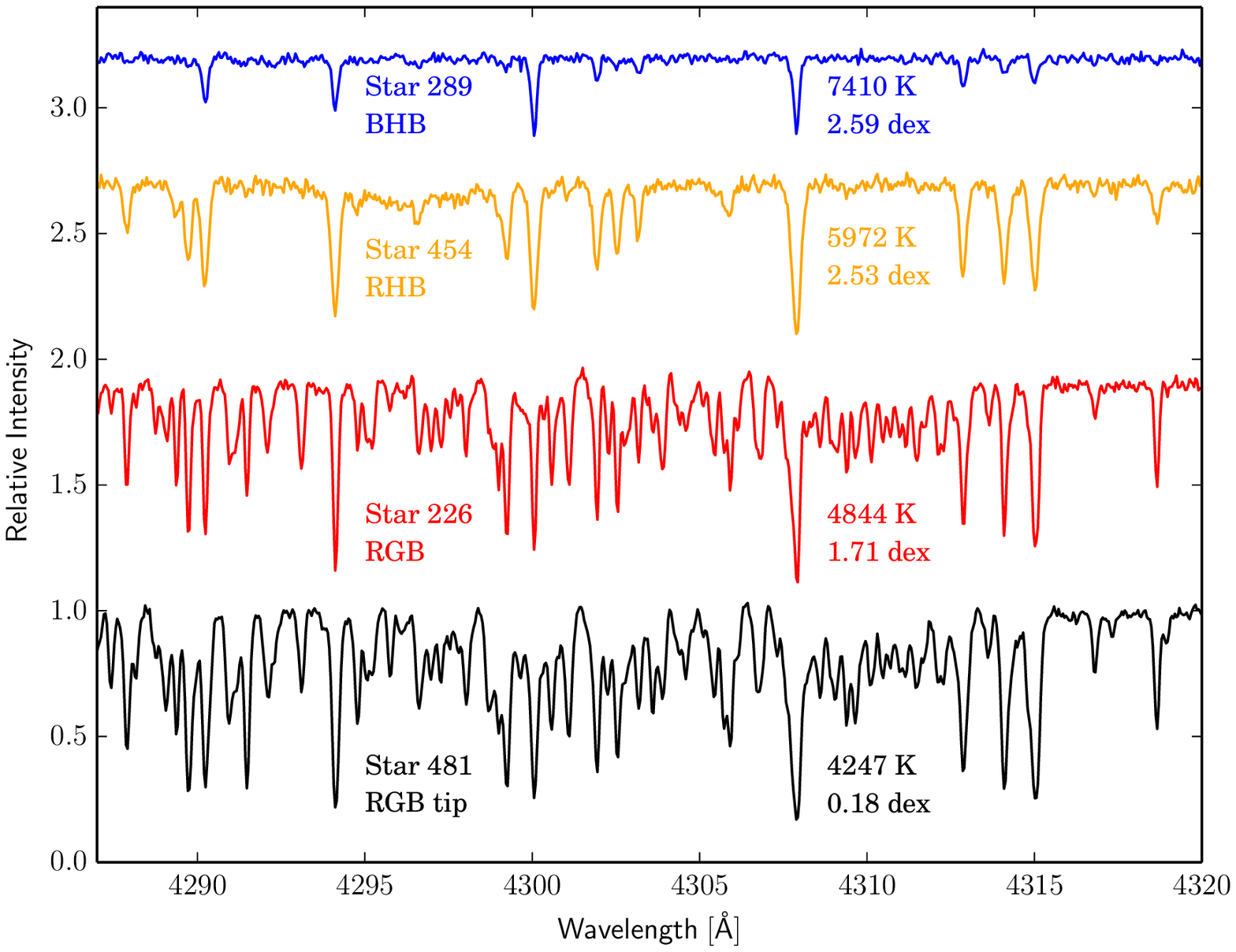}
\caption{
  \footnotesize{Sample spectra for all evolutionary stages in our sample. In addition to the name and evolutionary state of the star, 
  we also list \teff\ and \logg. This specific wavelength region covers the CH G-band. Measuring this band in the RHB and BHB is nearly 
  impossible due to their higher \teff\ and \logg\ values.}
}
\label{spectra}
\end{figure}

In this paper, we derive atmospheric parameters in a self-consistent fashion 
for RGB, RHB, and blue horizontal branch (BHB) stars in M68, which span about 4000 K in \teff\ and 
3.5 dex in \logg.
Figure~\ref{spectra}, which shows a small spectral region for 
all evolutionary stages in our sample, highlights the difficulties of a 
self-consistent analysis over such a large parameter space.
Several challenges, such as the breakdown of photometric temperature 
calibrations, as well as the unphysicality of certain spectroscopic 
methodology assumptions, have to be overcome.
After exploring photometric and spectroscopic methods to determine the 
values of \teff, \logg, \vmicro, and [Fe/H], we develop a hybrid 
atmospheric analysis strategy that appears to yield
reasonable parameters for the RGB, RHB, and BHB stars.
Using these parameters, 
we present a detailed abundance 
analysis, which will allow us to gain 
insight about the chemical evolution of M68.

\section{OBSERVATIONS AND REDUCTIONS\label{obsredEW}}
\begin{figure}
\epsscale{0.8}
\plotone{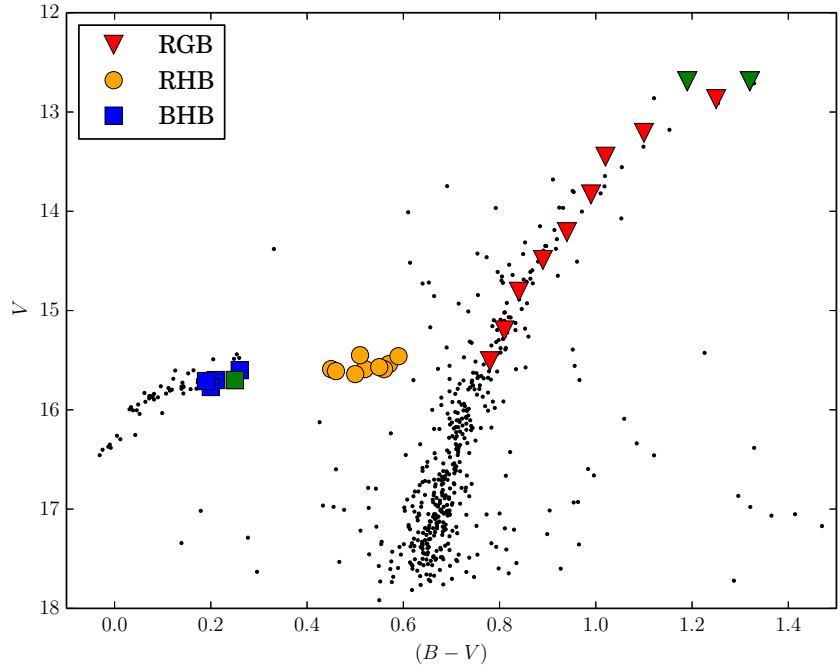}
\caption{
  \footnotesize{CMD of M68 using \cite{walker_bvi_1994} data. The markers and colors
  defined here for the RGB, RHB, and BHB will be used throughout the remaining plots
  in this paper. The green points represent non-standard stars. Please see \S\ref{isochrone_comp}
  for more details.}
}
\label{cmd_m68}
\end{figure}

We obtained high resolution spectra of 11 red giant branch, 
9 red horizontal branch, and 5 blue horizontal branch members 
of M68. 
All of our program stars were selected from the photometric survey 
of Wal94, whose $V$ and $B-V$ values are listed in
in Table~\ref{photometric_data}.
Figure~\ref{cmd_m68} shows the Walker color-magnitude diagram for
M68, using symbol shapes and colors to distinguish the RGB, RHB, and BHB stars
observed in this study.
The observed RGB members were selected to represent stars from the giant branch tip 
down to the luminosity of the horizontal branch, while RHB candidates 
were chosen to cover most of their evolutionary stage up to the red edge 
of the RR~Lyr gap ($B-V$~$\simeq$~0.45).
BHB candidates were chosen to be between the blue RR~Lyr edge ($B-V$~$\simeq$~0.25) 
and the \teff\ domain ($B-V$~$\gtrsim$~0.2) in which stellar atmospheric 
effects begin to distort the observed chemical compositions of 
stars (eg., \citealt{khalack_vertical_2010}, \citealt{behr_chemical_2003} and references 
therein).
Although M68 has more than 40 known RR~Lyr stars 
(\eg, \citealt{castellani_rr_2003}
and references therein), none of these were included in 
our study since spectrograph integration times would have been too 
long to acquire adequate data for these rapidly changing stars. 

Our spectra were gathered with the Magellan Inamori 
Kyocera Echelle (MIKE) spectrograph \citep{bernstein_mike:_2003}\footnote{
http://www.lco.cl/telescopes-information/magellan/instruments/mike}
of the LCO Magellan Clay 6.5 m telescope.
The spectrograph was configured with a 0.7$\arcsec$ entrance aperture that
yielded an ultimate resolving power of $R \equiv \lambda/\Delta\lambda$
$\sim$~40,000 for both the blue and red arms of instrument.
The useful spectral coverage of the blue arm was 3500$-$5000~\AA, 
and that of the red arm was 5000$-$9000~\AA. Scattered light and sky 
subtraction, as well as cosmic-ray filtering 
and flat field division were performed with software developed
by S.A. Shectman (2004, unpublished). Wavelength calibrations
were based on co-added hollow cathode Th-Ar spectra, obtained
before and after each observation. Other one-dimensional 
extractions were completed using the \textit{apall} package
of IRAF\footnote{IRAF is distributed by the NOAO, which is 
operated by the Association of Universities for Research in Astronomy, Inc., 
under cooperative agreement with the National Science Foundation.}. 

\section{LINE LISTS AND EQUIVALENT WIDTHS\label{lineEW}}
\subsection{Atomic Line List\label{atomic_line_list}} 
Table~\ref{line_list} lists the lines of atomic species and their associated
log(\textit{gf}) values employed in this work. 
Measured equivalent width ($EW$) values, as well as the sources of
our $gf$-values can be found in last column of this table.
We call attention to \ion{Fe}{2}, for which
\cite{melendez_both_2009} have proposed a renormalization of its transition
probabilities based on some laboratory $gf$'s and an inverted solar 
spectrum analysis. 
These renormalized $gf$ values are in general about 0.1~dex lower than 
those in the NIST Atomic Spectra Database\footnote{
NIST is the U.S. National Institute of Standards and Technology;
for the atomic line database see 
http://physics.nist.gov/PhysRefData/ASD/lines\_form.html}, which were 
employed by us.
If we had adopted the \citeauthor{melendez_both_2009} results here,
our \ion{Fe}{2}-based abundances would be larger by about 0.1~dex,
which, in turn, would have decreased derived gravities by 
about 0.2~dex. We will return to this point later.

\subsection{\textbf{Equivalent Width Measurements}\label{ewmeasure}}

The equivalent width (\textit{EW}) measurements were done in a 
semi-automated manner using an IDL code ({\sf EW.pro}) initially 
described in \cite{roederer_characterizing_2010} and further developed 
by \cite{brugamyer_silicon_2011}. 
This code allows the user to visually inspect either a Gaussian or 
Voigt $\chi^{2}$ minimization fit for each line, ensuring that any 
obviously blended or any otherwise undesirable line will not be measured. 
Additionally, the user can adjust the continuum which reduces any 
error possibly introduced by faulty normalization of the spectra. 
Several lines were picked out at random from all stars 
and re-measured using the {\sf SPECTRE} code 
\citep{fitzpatrick_software_1987}\footnote{
Available at http://www.as.utexas.edu/\textasciitilde chris/spectre.html}.
Four members of our sample 
(RGB stars 256, 472 and RHB stars 403, 458)
were analyzed entirely using {\sf SPECTRE}. 
The \ew s obtained in this fashion were then compared to those derived 
using the IDL code. 
On average, the difference between the \textit{EW} values derived using 
{\sf EW.pro} and those derived using {\sf SPECTRE} are 
$\Delta_{EW} = -1.18 \ (\sigma$ = 3.18) m\AA. 
We therefore regard the differences in measured \textit{EW} as negligible. 

The \textit{S/N}, which is a function of wavelength and 
\teff\ of our programs stars, directly influenced 
the \textit{EW} limitations for lines used in our analysis.
A set of final \textit{S/N} estimates for all of our stars can be found 
in Table \ref{atm_params}. All \textit{S/N} values given in this Table
were estimated using the \textit{no} routine of {\sf SPECTRE} at 
around 6600 \AA\ in each star. 
For most stars, the lower \ew\ cutoff was $\sim$ 10 m\AA, 
while the upper limit was set at $\sim$ 150 m\AA, depending on evolutionary 
state and line species.
These limits, especially the lower cutoff, were not applied to star 334, 
as its \textit{S/N} is about 70\%  
lower than the average of our sample. For this star, any line that could 
be measured with a healthy degree of certainty, except the obviously 
saturated ones, was used.

\section{MODEL ATMOSPHERIC PARAMETERS\label{models}}
\subsection{Initial Parameters\label{initial}}

General information about M68, relevant to determining atmospheric 
parameters of our stars, can be found in Table \ref{fun_params}. 
Special attention is called to the assumed distance modulus, 
$(m-M)_{V}$, and the reddening, $E(B-V)$, as they are of importance in 
determining photometric \teff \ and \logg \ values.

\begin{figure}
\epsscale{0.8}
\plotone{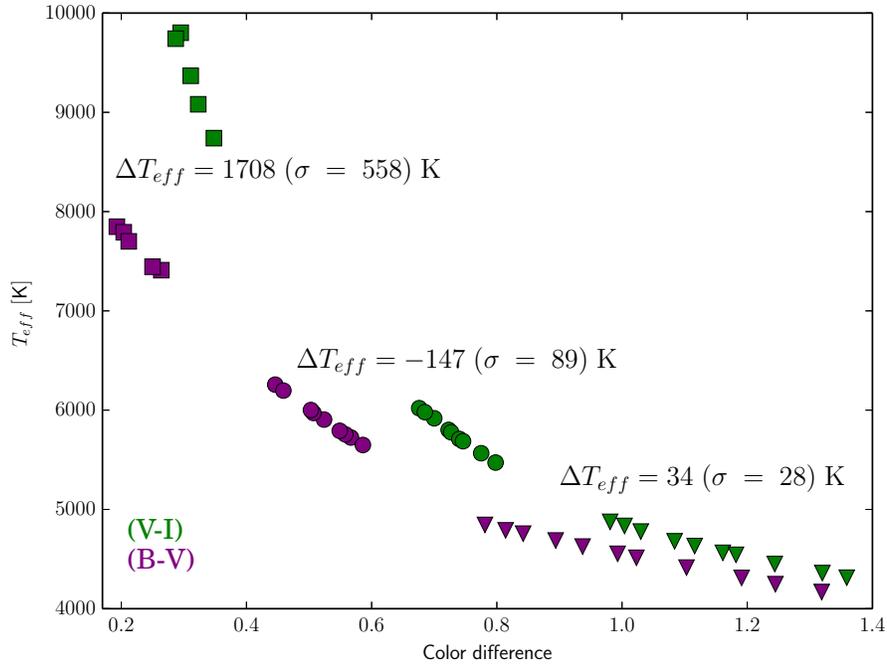}
\caption{
  \footnotesize{Here we show the implied \teff\ values for each of the color indices. For the purposes of this
  figure, $\Delta \teff\equiv\teff_{(V-I)} - \teff_{(B-V)}$. As before, RGBs are represented as triangles, RHBs as circles, and
  BHBs as squares.}
}
\label{color_comp_teff}
\end{figure}

Initial atmospheric parameters were determined from \textit{BVI} 
photometry obtained by Wal94 (see Table \ref{photometric_data}). 
To convert the given fluxes to \teff \ values, the \textit{(B-V)} and 
\textit{(V-$I_{C}$)} IRFM (infrared flux method) calibrations of 
\cite{ramirez_effective_2005} were used.  
Although there is a more recent calibration available 
(\citealt{casagrande_absolutely_2010}), only \cite{ramirez_effective_2005} 
include a calibration for giant stars. 
Figure \ref{color_comp_teff} compares the \teff\ values obtained from
\textit{(B-V)} and \textit{(V-$I_{C}$)} colors.
In the RGB, these two indices give approximately the same answer, while 
they start to diverge in the RHB and BHB. This behavior can be explained by the lack
of calibration stars at \textit{(V-$I_{C}$)} $<$ 0.6. 
Additionally, the difference between the \textit{V} and $I_{C}$ fluxes 
becomes insensitive to changes in temperature at 
\textit{(V-$I_{C}$)} $<$ 0.65, since the bandpasses of the respective
filters are now in the 
temperature insensitive tail of the blackbody distribution.
For these reasons, we chose \textit{(B-V)} to be the sole 
color index in determining photometric temperatures.

Initial surface gravity values were derived using the standard formula: 
\begin{displaymath}                                               
\logg_{\star} = 0.4 (M_{\rm V\star} + BC - M_{\rm Bol\sun}) + \logg_{\sun} +    
4 {\rm log} \left(\frac{\teff_{\star}}{\teff_{\sun}}\right) + {\rm log} \left(\frac{{\it M}_{\star}}{{\it M}_{\sun}}\right).
\end{displaymath} 
For the solar values, $\ M_{\rm Bol\sun}$\ = 4.75, 
\teff$_{\sun}$ = 5777 K,
and \logg$_{\sun}$ = 4.44~km~s$^{-1}$ were assumed. 
The assumed stellar mass in the above equation was estimated to be
$M_{\star} \approx $\ 0.7$\ M_{\sun}$, a result obtained from 
a isochrone calculation with M68 metallicity and age values
(see \S\ref{isochrone_comp} for more details).
Since $\logg_{\star}$ varies linearly with log($M_{\star}$), an accurate 
value of stellar mass was not needed in this calculation.
The bolometric correction (\textit{BC}) for each star was calculated from 
the calibration given in \cite{alonso_effective_1999}. 
This calibration, however, does not hold for stars with \teff $\ \geq $\ 6300~K.
For stars exceeding this temperature, 
Figure~3 of \cite{flower_transformations_1996} was used to obtain an 
estimate of the stellar \textit{BC}. 
\teff\ and \logg\ values obtained in this fashion will
be referred to as PHOT in texts and figures throughout the 
rest of this paper. 

Initial microturbulent velocities (\vmicro) were estimated to be 1.2~\kmsec\
for all RGB stars in our sample. 
For RHB stars, calibrations provided in \cite{gratton_abundances_1996} 
were used. 
Initial BHB \vmicro \ estimates were obtained by calculating a temperature 
dependent linear fit of a previous BHB study (\citealt{for_chemical_2010}), 
and applying the resulting calibration to our stars.
Lastly, an initial atmospheric metallicity of 
[Fe/H] = $-$2.23 (from the 2010 edition\footnote{
http://www.physics.mcmaster.ca/\textasciitilde harris/mwgc.dat}
of \citealt{harris_catalog_1996}) was assumed.

\subsection{Final Parameters\label{finalparams}}

We employed three different methodologies to converge on our final 
set of parameters. In addition of the purely photometric PHOT
parameter set defined above, we also derived purely spectroscopic
parameters. These will be referred to as SPEC.
Our final adopted set of parameters, which are
a combination of photometry and spectroscopy, are 
called COMB. 
These designations will be used throughout the rest of this paper, including
all plots.
Abundances for the various approaches were calculated using the 
latest version of {\sf MOOG} (\citealt{sneden_nitrogen_1973})\footnote{
Available at http://www.as.utexas.edu/\textasciitilde chris/moog.html}, except 
in the case of the RGB stars. 
In their temperature/gravity/metallicity regime, the major electron donors
for the $H^{-}$ continuous opacity are $Fe$ and
$\alpha$ elements.
Since these elements are very deficient in metal poor stars such as
M68 RGB members, the $H^{-}$ opacity
decreases significantly, and scattering processes
become important in the blue-UV spectral regions.
Therefore, we employed a {\sf MOOG} version incorporating
Thompson scattering (see Sob11 for more details) for this
subgroup of our sample. 
[X/H] abundances were calculated using the solar abundance
recommendations of \cite{asplund_chemical_2009}. 
Our model atmospheres were interpolated from ATLAS9 
$\alpha$-enhanced
opacity distribution function model grids (\citealt{castelli_new_2003})
using software developed by Andy McWilliams and Inese Ivans.

To obtain the final atmospheric parameters for our stars, we decided
to employ the analytical tools of the `classical' spectroscopic approach, 
but deviated somewhat from its exact prescriptions.
 Specifically, we adopted photometric temperatures and then used the well 
known plots of individual \ion{Fe}{1} line abundances as a function of 
excitation potential (EP) and as a function of reduced equivalent width,
log(RW) $\equiv$ log(EW/$\lambda$) to estimate 
microturbulent velocities. 
These photometric \teff \ values were generally higher than the 
spectroscopic ones (see \S\ref{comparison_photo_spect}
for more details), and
an undesirable positive trend in the EP plot described above was obtained. 
This trend was mitigated by adjusting \vmicro \ until acceptable 
correlations for both the EP and log(RW) plots were reached. 
Positive/negative changes in \vmicro\ generally decrease/increase the 
abundances derived from strong lines;
weak lines are generally not affected by changes in \vmicro.

The \logg \ values of our final approach were obtained by 
requiring equality between the abundances of [\ion{Fe}{1}/H] and 
[\ion{Fe}{2}/H]. This parameter determination process invariably changed the 
stellar metallicity, and therefore also implied photometric 
temperatures of the individual stars, forcing us to 
repeat the above process with these altered photometric \teff\ values.
Metallicities resulting from this second iteration differed very little 
from those of the first iteration, eliminating the need for a third iteration. Lastly,
the metallicity of our atmosphere was equalled to $\langle$[Fe/H]$\rangle$ of each star.
Atmospheric parameters obtained from the above methodology 
can be found in Table~\ref{atm_params} under the COMB heading. 
Unless indicated otherwise, any further mention of atmospheric parameters will refer to these 
COMB values.

Before highlighting two main successes of this approach, we review the 
methodology of the `classical' spectroscopic approach since we compare 
results of our final approach to those of the spectroscopic method below. 
Spectroscopic parameters (designated SPEC) were derived by: 
(1) requiring no trend with solar normalized abundances of \ion{Fe}{1} 
and \ion{Fe}{2} with excitation potentials (EPs) of different lines 
(giving \teff);
(2) forcing ionization balance between \ion{Fe}{1} and \ion{Fe}{2} 
(giving \logg);
(3) demanding a correlation coefficient smaller than 0.01 between solar 
normalized line abundances and the logarithm of the reduced equivalent 
widths (log(RW)) (giving \vmicro): and
(4) setting the atmospheric [Fe/H] equal to the resulting average 
normalized stellar iron abundance.

The first motivation for our COMB approach
is revealed in Figure~\ref{ionization_comp}, where we
show trends of Fe and Ti 
abundances with \teff. 
By design, the abundances of the neutral and ionized lines Fe lines
agree, leading us to only show one Fe point per star.
\ion{Ti}{1} and \ion{Ti}{2} abundances are shown with
separate symbols. For the purposes of this discussion, we will ignore
the BHB stars, as they present special challenges. 
See \S\ref{bhb} for more details. 
Since our targets are all confirmed members
of M68, the \teff-[Fe/H] trends displayed in 
the upper panel of Figure~\ref{ionization_comp} (SPEC parameters)
are unphysical and unexpected.
Metallicities obtained from the adopted COMB (photometric, bottom panel) \teff\ still show 
a positive trend with increasing temperatures, but it is much less pronounced. 
Additionally, the COMB [Fe/H] values agree better with current metallicity estimates of M68.
We attempted to eliminate the \teff-[Fe/H] trend in the RHB by 
using microturbulent velocities to correct for 
any differences between our stars. This approach, however,
led to \vmicro \ values ranging from 
3.0 \kmsec \ to 16.0 \kmsec,
far too high for RHB stars.

Perhaps the more important reason for adopting our final approach
can be seen in Figure~\ref{logg_comp_rgb_rhb},  
where we plot 
$\Delta$\logg \ ($\equiv \logg_{PHOT} - \logg_{comparison \ value}$) versus
\teff \ for the RGB and RHB evolutionary stages. 
The top panels of each column compare spectroscopic and 
photometric \logg \ values from previous studies,
while the middle and bottom panels compare our SPEC and COMB \logg\ values to 
the PHOT ones (see \S\ref{initial} for more details about 
photometric \logg\ values).
Clearly, the differences between the purely photometric and our 
COMB \logg \ values, which are displayed on the y-axis in Figure
\ref{logg_comp_rgb_rhb}, are smaller than those of any other approach and
exhibit little to no trend with increasing \teff. 
These results will be discussed in more detail 
in \S\ref{comparison_photo_spect}.

As a final note, we call attention to stars 117, 160 (RGBs) 
and 324 (BHB), for which we were unable to derive atmospheric parameters
using either the SPEC or COMB approaches outlined above. 
Lee05, who also analyzed stars 117 and 160, believe that star 117 is an 
AGB rather than a RGB star. 
However, we suggest that both 117 and 160 might be
extreme examples of RGB tip stars, which are inherently difficult to analyze 
using standard spectroscopic methods.
For these two stars, the adopted \teff \ and \logg \ were obtained using the 
the purely photometric approach described in \S\ref{initial}. 
For \vmicro \ and [Fe/H], the average of all other RGB stars was chosen. 
Despite these differences in assigning atmospheric parameters, the 
resulting abundance patterns of these stars are in good agreement 
with other stars in our sample. 
If these stars are, in fact, AGB stars, 
our assumptions about the their atmospheric structures
could be false, explaining our analytical difficulties.
Star 324 (BHB) exhibited an unusually low \textit{S/N}, which, in 
combination with it being a BHB star, resulted in less than 20 measurable 
Fe lines, making a spectroscopic analysis very difficult. For this star, 
we performed 4 iterations of the COMB approach and then 
adopted the resulting parameters.

\subsection{Further Motivations for Our Atmospheric Parameter 
Derivation Approach}\label{comparison_photo_spect}
In the previous section, we highlighted two reasons
for adopting our atmospheric parameter derivation methodology. 
In addition to quantifying these advantages in this section, 
we will also contrast our COMB parameters to literature values in 
the following order: \teff, \logg, \vmicro, and finally [Fe/H].

Figure~\ref{teff_comp_rgb_rhb} compares the COMB and SPEC \teff\ values
listed in Table~\ref{atm_params} for the RGBs and RHBs. 
\teff\ values we derived from photometry are systematically higher 
than spectroscopic ones: 
$\langle \Delta \teff\rangle$ = $\langle \teff_{PHOT} - \teff_{SPEC} \rangle$ 
= 159~($\sigma$~=~80)~K for our study.
Lee05 found a similar systematic upward shift of photometric temperatures, 
with their data giving $\langle \Delta \teff\rangle$ = 100~($\sigma$~=~61)~K. 
For the two stars shared by the two studies (117 and 160), differences between
their photometric and our COMB \teff value are 60~K (star 117) 
and 32~K (star 160).
These small differences are most likely caused by Lee05's usage of slightly
different IRFM calibrations (\citealt{alonso_effective_1999}), 
as well as different $(m-M)_{V}$ and $E(B-V)$ values.
Overall, the results of the two studies are comparable.

For RHB stars in M68, there exists no previous 
high-resolution spectroscopic literature reference.
We will therefore compare our results to the M15 RHB results of Sob11. 
The average offset between the photometric and spectroscopic \teff \ 
values of our study is  $\langle \Delta \teff\rangle = 321 \ (\sigma~=~146$) K. 
Sob11 found a much smaller average difference of 
$\langle \Delta \teff\rangle = 51 \ (\sigma~=~272$) K. 
The constant offset between the photometric and spectroscopic \teff \ values 
of our study (see Figure~\ref{teff_comp_rgb_rhb}) suggests that perhaps 
our adopted reddening value was a bit too high, causing hotter photometric 
temperatures. 
Unfortunately, there are no recent M68 RHB studies available that allow us to 
further explore this difference.

A far greater discrepancy between photometric and spectroscopic parameters 
is present in derived \logg \ values. 
The RGB side of Figure \ref{logg_comp_rgb_rhb} clearly shows a trend 
between adopted final \teff \ and $\Delta$\logg \ values 
(definition given above).
In the upper panel of this figure, we have included the Lee05 
$\Delta$\logg \ values, which exhibit a slight upward trend 
of $\Delta \logg$ with increasing $(B-V)$ temperature. 
Our data (middle and bottom panel) shows a similar trend.
In the specific case of star 160, Lee05 derive a spectroscopic \logg \ 
of 0.0 dex, whereas we were unable to derive a spectroscopic \logg \ value. 
For the photometric \logg, Lee05 derive a value of 0.7 dex, close
to our value of 0.65 dex. Given difference in adopted distance 
modulus and reddening value, this discrepancy is not serious. 
For star 117, Lee05 derive a spectroscopic \logg \ of 0.3 dex.  
We were again unable to derive a spectroscopic \logg. 
The photometric \logg \ values are very close, with ours being 0.75 
and Lee05 giving a value of 0.8. 
The average difference between the photometric and spectroscopic \logg\ 
obtained by Lee05 is $\Delta \logg$ = $0.67 \ (\sigma=0.10)$, while the difference 
between our photometric and COMB \logg\ values is only
$\Delta \logg$ = $0.47\ (\sigma~=~0.18)$. 

The purely photometric study of \citeauthor{carretta_na-o_2009} (2009, hereafter Car09)
also has two RGBs in common with our study, 57 and 79.
Since Car09 employed nearly identical $(m-M)_{V}$ and $E(B-V)$ values,
differences between our COMB and their final parameters are:
-43 K in \teff \ and -0.31 in \logg\ for star 57, while star 79 exhibits differences
of -29 K in \teff \ and -0.44 in \logg.
Given the difficulties of analyzing extreme RGB tip stars such as 160 and 117, 
we lend more weight to the differences between our study and Car09, which, as 
just demonstrated, are not severe.

The $\Delta$\logg \ comparisons for our RHB stars are shown on the 
right hand side of Figure~\ref{logg_comp_rgb_rhb}. 
The top panel shows the Sob11 M15 data, which seems to exhibit
a fairly strong $\Delta$\logg-\teff\ trend if compared to our data (middle and
bottom panel).
Our COMB $\Delta$\logg \ values give
$\langle\Delta\logg\rangle$ = $0.22\ (\sigma~=~0.09)$, while 
$\langle\Delta\logg\rangle$ = $0.72 \ (\sigma~=~0.41)$ for our SPEC parameters. 
The Sob11 data exhibit $\langle\Delta\logg\rangle$ = $0.36 \ (\sigma~=~0.38)$. 
Clearly, the average differences (dashed lines) as well as $\Delta$\logg-\teff\ trends
are minimized for our COMB parameters in both the RGB and RHB stars.

As alluded to in \S\ref{atomic_line_list}, all of our derived \logg\ values would be $\sim$ 0.2 dex lower
if we had adopted the \cite{melendez_both_2009} based \ion{Fe}{2} log($gf$) values, instead of the
NIST values. Hence, the $\Delta\logg$ values in  Figure~\ref{logg_comp_rgb_rhb} 
would be enhanced by about 0.2 dex. 
However, even with such an enhanced difference, the current \teff \ trends would
still exist. An additional effect that we have neglected so far
is the temperature dependence of the physical \logg \ equation given in \S\ref{initial}.
To quantify this effect, we adopted our SPEC temperatures to determine 
photometric \logg \ values, which lowered all of our physical \logg \ values
by $\sim$ 0.1 dex. Since this value is much lower than our typical \logg \
uncertainties (see Table~\ref{uncertainties}), we can safely disregard the \teff \ dependence of the photometric
\logg. In fact, \teff \ changes of 600 K or more are needed to reproduce
\logg \ shifts equivalent to our derived \logg \ uncertainties. Please see \S\ref{errors}
for more details.

Figure \ref{vt_comparison} shows a comparison of our derived \vmicro \ to 
those of previous studies and the calibrations of 
\cite{gratton_abundances_1996}. 
In particular, we used the theoretical PARSEC \teff\ and \logg\ values, 
applied these to the RGB and RHB \vmicro\ calibrations of 
\cite{gratton_abundances_1996}, and plotted the results as thick dashed 
lines in Figure \ref{vt_comparison}
with the label `PARSEC \vmicro'. Our RGB results agree well with those of previous studies (\citealt{cayrel_first_2004})
and the RGB \vmicro \ calibrations in the high temperature end. At low temperatures (\teff \ $\approx$ 4200 K), 
our microturbulent velocities seem to deviate from 
the empirical fits of \cite{gratton_abundances_1996}. 
However, since stars in this temperature regime are difficult to analyze, 
we do not lend much weight to this difference. Individual comparisons with previous studies of
Lee05 and Car09 are not possible since Car09, being a purely photometric study, do not
give \vmicro \ values and for the two stars shared with Lee05 (116 and 170), we adopted
average \vmicro \ values. See \S\ref{finalparams} for more details.
Our RHB stars agree well with those of \cite{for_chemical_2010}. 
The PARSEC trend of decreasing \vmicro \ with decreasing \teff \ is also 
followed by both our sample and that of \cite{for_chemical_2010}. 
Given this good agreement, we regard our RHB \vmicro \ values as satisfactory.
We defer discussion of BHB \vmicro\ values to the next section.

Lastly, we compare the resulting [Fe/H] values of our 
COMB approach to those of previous studies,
which include Car09, Lee05, 
\cite{behr_chemical_2003}, and \cite{harris_catalog_1996}. The difference 
between our study, which resulted in $\mathrm{[Fe/H]_{M68}}$ = -2.41, 
and that of Car09 is $\Delta\mathrm{[Fe/H]}$\footnote{$\Delta\mathrm{[Fe/H]}$ = 
$\mathrm{[Fe/H]_{previous \ study}}$
- $\mathrm{[Fe/H]_{our \ study}}$} = 0.15, while for \cite{harris_catalog_1996},
$\Delta\mathrm{[Fe/H]}$ = 0.18. Lee05 doesn't give a definite [Fe/H] value, 
but upon averaging their [\ion{Fe}{1}/H] and [\ion{Fe}{2}/H] values, we 
obtain $\Delta\mathrm{[Fe/H]}$ = 0.07. 
Differences between our study and that of \cite{behr_chemical_2003} yield
$\Delta\mathrm{[Fe/H]}$ = 0.13.
We conclude that there are no
significant metallicity differences between our and previous studies.

All of the atmospheric parameters described above were
derived without considering Fe NLTE effects. Our assumption
of ionization balance, which we used to obtain \logg \ values, 
has been identified as potentially leading to incorrect atmospheric
parameters (i.e. \citealt{fabrizio_carina_2012} and \citealt{bergemann_non-lte_2012}).
To quantify the severity of these effects in our sample
stars, we used Figure 4 of \cite{bergemann_non-lte_2012} to estimate
NLTE corrections for our derived \logg \ values. Unfortunately, we were only able to do this 
for some of our RGBs (all except 117, 160, 450, 472, and 481). 
We are currently unaware of any published NLTE calculations
for RHB and BHB stars. After applying the NLTE corrections 
to our RGBs, we found that our \logg \ values were raised
by approximately 0.3 dex, which, in turn, would essentially 
eradicate the difference between our PHOT and COMB \logg \ 
values shown in Figure~\ref{logg_comp_rgb_rhb}. However, it would
also lead to greater differences between our atmospheric parameters
and PARSEC isochrones. This will be discussed in further detail in 
\S\ref{isochrone_comp}. Since NLTE Fe calculations 
are only available for a small selection of RGBs in our sample, we 
decided to ignore these effects for our analysis (see
\S\ref{abunds} for a discussion of the effects of this
choice on derived abundances).  A future effort that considers
NLTE effects over the whole parameter range of 
evolved stars in this cluster is welcome.

\subsection{Challenges of the BHB stars}\label{bhb}
The BHB stars in our sample suffer from photometric 
and spectroscopic deficiencies, which need to be discussed before
proceeding to analyze their abundances. 
Photometric difficulties include the following:

(I) the \cite{ramirez_effective_2005} calibrations for our metallicities 
become unreliable at approximately 7000K due to a lack of reliable calibration 
stars

(II) at \teff $\ \geq $\ 7000K, the peak of the stellar blackbody curve 
lies at shorter wavelengths than 
the centers of the \textit{B} ($\sim$4400 \AA) and 
\textit{V} ($\sim$5500 \AA) bandpasses.
Therefore, fluxes are now being measured in the temperature-insensitive 
Rayleigh-Jeans tail of the Planck distribution, resulting in \teff \ values 
that are extremely insensitive to changes in either the \textit{B} or 
\textit{V} flux.
Other photometric fluxes (mainly \textit{K}) are available for some of 
our stars, but only \textit{(B-V)} is available for the entire sample. 
Moreover, the offset between the flux peaks of the BHB stars and \textit{V-K} 
bandpasses is worse than for \textit{B-V}.
We therefore did not consider
any photometric \teff \ or \logg \ values derived from \textit{(V-K)}.

The spectroscopic temperatures are also afflicted by certain weaknesses. 
Due to the higher temperatures of the BHB stars, many lines, especially 
those with high excitation potentials, 
are not present in their stellar spectra. 
The reason for this behavior is that the strength of 
absorption lines is dependent on both the line and continuous absorption. 
The continuous absorption, mainly due to $H^{-}$ (and neutral $H$
in the violet and near-UV spectral regions), increases sharply with
temperature and is much larger in BHB stars than in RHB and RGB stars.
In our BHB stars, we therefore see a severe drop-off in the availability of weak high EP lines. 
The absence of these lines introduces a bias when determining the 
atmospheric parameters purely from spectroscopic line analysis.
Furthermore, the increased temperatures also eliminated one of our 
atmospheric parameter diagnostic elements, \ion{Ti}{1}. 
The absorption lines of this species are not strong
even in RHB stars, and become undetectable in the BHB stars.
Given the difficulties in both photometric and spectroscopic 
approaches, we advise the reader to treat 
all of the atmospheric parameters of the BHBs with caution.
A more involved treatment of M68 BHB stars can be found in 
\cite{behr_chemical_2003}.

Our simple analysis of the BHBs, however, seems to produce \vmicro \ values that are
somewhat comparable to those of previous studies. 
Figure \ref{vt_comparison} 
shows a comparison of our data to a linear fit (thin dotted line) of the RR-Lyrae of
\cite{for_chemical_2011} and \cite{govea_chemical_2014}
and the BHB of \cite{for_chemical_2010}.
For this fit, we have excluded BHB stars with $v \ sin \ i > $ 15 km/s. 
It has been suggested by \cite{govea_chemical_2014} that BHB 
stars with larger rotational velocities suffer from abnormally large 
\vmicro\ values; see that study for more details on this point.
Our values fit very well with all of the previous data and we therefore regard our
microturbulent velocities as satisfactory.

\subsection{Comparison of Final Atmospheric Parameters with Isochrones\label{isochrone_comp}}
PARSEC isochrones (\citealt{bressan_parsec:_2012}) exhibit several attributes 
which ultimately led us to choose them as the master isochrones for this 
project. Most importantly, PARSEC isochrones allow for the 
computation of isochrones for arbitrary input metallicities and ages.
This is due to the usage of two different types of opacities during the 
calculations. For the low temperature regime, opacities are obtained 
from the \AE SOPUS code (\citealt{marigo_chemical_2001}), while high 
temperature opacities are calculated using OPAL 1996 
(\citealt{iglesias_updated_1996}) data. 
Additionally, the latest version of the freely available 
FREEEOS\footnote{http://freeeos.sourceforge.net/} is used to 
derive the equations of state. Another very important aspect
of the PARSEC isochrones show the \teff \ - \logg \ relationship
beyond the tip of the RGB, a usual stopping point for 
other sets of calculations. 
All of these improvements over 
older sets of isochrones should provide a good estimate of the 
relationship between surface gravities and effective temperatures for 
different evolutionary states in M68. 

Figure~\ref{isochrone} compares a PARSEC isochrone calculated 
assuming an age of 11.2 Gyr and a 
metallicity of Z = 1.178\e{-4} (both consistent with current M68 estimates)
with our COMB (filled symbols, usual colors) and SPEC (unfilled symbols) parameters.
Overall, our final data matches the PARSEC isochrone much better, 
providing additional motivation for our COMB approach.
We also included NLTE corrected parameters
in Figure~\ref{isochrone} (black triangles). As mentioned in \S\ref{comparison_photo_spect}, 
these values show greater discrepancies with the isochrones. 
To close this now bigger gap between the NLTE parameters and PARSEC isochrones, 
the metallicity of our evolutionary calculations would have to be increased by $\sim$ 0.4 dex, leading
to an inferred $\mathrm{[Fe/H]_{M68}} \sim$ -1.80 dex, a value completely
at odds with previous literature studies of this cluster. 
We also note that isochrone calculations are nearly age independent
 at M68 metallicities, and thus this difference cannot be accounted
for by an adjustment of this input parameter.

A comparison of the Wal94 photometry data to this isochrone in color 
space is shown in Figure \ref{isochrone_color}. 
In both of these figures, we have also included the latest version of 
BaSTI isochrones (\citealt{pietrinferni_large_2004}) with the age and
metallicity of M68. 
In the \teff \ - \ \logg \ plane, the difference between the two isochrones 
is negligible, while in the color plane, the two isochrones give 
very different answers. The better fit of our data to the PARSEC 
isochrone tracks in the color plane provides further 
motivations for using these calculations.

\subsection{Uncertainties\label{errors}}
Instead of deriving uncertainties for all stars in our sample,
we chose representative members 
of four different evolutionary stages in our sample: lower RGB (172), 
RGB tip (472), RHB (36) and BHB (337). 
To obtain the \teff \ uncertainty, we first calculated the errors in the 
photometric temperature by considering the color uncertainties given in Wal94. 
For all of the evolutionary stages, this amounts to $\sigma_{PHOT}\simeq$ 20K 
for the RGB stars and  $\sigma_{PHOT}\simeq$ 70K for RHB stars.
Since errors stemming from the scarcity of available measured lines 
dominate in the BHBs, this part of the error analysis was not performed for this evolutionary 
stage.
In addition to photometric errors, we also considered errors in the 
reddening, which amounted to  $\sigma_{E(B-V)}\simeq$ 10K for the RGBs 
and  $\sigma_{E(B-V)}\simeq$ 45K for RHBs. 
The measurement uncertainty contribution to our total $\sigma_{\mathrm{\teff}}$
was simulated by 
adjusting the temperature of stars until we achieved a shift in \ion{Fe}{1} 
abundances equal to the standard deviation of the abundances implied by the 
stellar \ion{Fe}{1} lines.
The errors from the photometry and measurements were added in quadrature 
to give the final \teff \ uncertainty. 
Uncertainty in \logg \ was then determined by applying the upper and lower 
values of \teff \ to our stars and then again requiring ionization balance. 
For \vmicro \ uncertainties, we repeated the process that we used to 
determine final atmospheric parameters (see Sec. \ref{finalparams}), 
but this time applying the upper and lower temperature limits. 
The final uncertainties determined in this manner can be found in 
Table \ref{uncertainties}. 
We did not consider atmospheric metallicities in our 
error analysis, as they have a negligible effect on total
parameter errors.

The differences in uncertainties between the evolutionary stages is easily 
explained. Atmospheres of RGB stars are much more sensitive to \teff \ and \logg \ 
changes than the more evolved stars due to the greater number of observed lines
and greater range of excitation potentials, ionization state and log(\textit{gf})
values. As \teff \ and \logg \ increase, we see a drop off in the number of observed lines
and therefore less sensitivity to changes in atmospheric parameters. Hence, our RHB 
have larger uncertainties than our RGBs, but smaller uncertainties  than the BHBs, just
as expected.

\section{ABUNDANCE ANALYSIS\label{abunds}}
In this section we present the results of our abundance analysis. 
Unlike the discussion of the atmospheric parameters, we will discuss these 
not individually by evolutionary stage, but rather attempt to give an 
overview of post main sequence stars in M68. 
All of the following abundances are derived using the COMB parameters 
given in Table~\ref{atm_params}. Table~\ref{individual_abunds} gives 
average abundances in standard [X/Fe] fashion for each star in our
sample, while Table~\ref{evolution_abund} summarizes the average abundances 
for each evolutionary stage and M68 as a whole. The first two lines in 
these tables give [Fe/H] for the corresponding star/evolutionary 
stage. We note here that the \ion{Fe}{1} and \ion{Fe}{2} of stars
117 and 160 do not agree. The reason for this non-agreement
 is explained in \S\ref{finalparams}.

For completeness, we will now briefly discuss the sensitivites of
 our abundances to the adopted atmospheric parameters, which we determined
 by deriving abundances using the SPEC and PHOT parameters given in
 Table~\ref{atm_params} for stars 472, 172, 36, and 337
 (stars used to derive atmospheric parameter uncertainties). 
 Since we did not derive PHOT \vmicro \ and [Fe/H] values, we adopted those of the COMB
 approach for the stars listed above. 
 This is justified by repeating the analysis described in \S\ref{finalparams}, 
 but this time fixing \logg\ to the photometric value, which showed
 that the resulting PHOT \vmicro \ and [Fe/H] values differ very little from
 those of the COMB approach.
The derived absolute abundances shifted as a result of the adoption of the
 PHOT and SPEC parameters, but the metallicity, which we use to normalize
 our abundances, changed by approximately the same amount (see Table~\ref{atm_params}). 
 Therefore, the resulting normalized abundances differed from the 
 COMB abundances by less than the
 abundance uncertainties discussed in \S\ref{uncert}
 and listed in Table~\ref{uncertainties}.
 We thus anticipate 
 no serious changes to our derived abundances
 due to the adoption
 of either the PHOT or SPEC parameters.
 
A summary of the element ratios ([X/Fe]) for $8 \leq Z \leq 70$ in
box-plot form can be found in Figures~\ref{light_elements_1},~\ref{light_elements_2},
and~\ref{heavy_elements}. Unfortunately, 
the quality of our spectra did not allow us to make any inferences about 
differences in He abundances between M15 and M68 (see \S\ref{intro}
for more details).
As before, red data represents the RGB, yellow data represents
the RHB, and the blue data shows the BHBs. We have also 
added data from previous studies, which is shown as grey boxes. 
In these box plots, the mean elemental abundance is represented by the central
bar, while the upper and lower edges of each box give the first (lower edge)
and third (upper edge) quartile of the plotted data. The whiskers
constitute a $\sim 3\sigma$ boundary. 
Outlier data beyond the whiskers is depicted as solid symbols with corresponding colors.
Table~\ref{abund_sources} gives a summary of previous studies used for comparison
in Figures~\ref{light_elements_1},~\ref{light_elements_2}, and~
\ref{heavy_elements}.
A list of lines, as well as the method used to derive the abundances 
(EW or synthesis) are given in Table \ref{line_list}.

\subsection{Light and Fe-Group Elements\label{light_ele}}

\textit{Carbon and Nitrogen}:
Due to the low metallicity of M68 and the \textit{S/N} 
limitations of our spectra, determining \ion{C}{1} and \ion{N}{1} abundances for our sample 
stars was very difficult. 
After co-adding the spectra of our four coolest RGB stars,
we were 
able to estimate $\langle$\ciso$\rangle$ $\simeq$ 5$-$7 for the coolest
giants in M68. The quality of our stellar spectra precluded 
determination of values for individual stars. 
Better quality data are needed to make a more accurate carbon isotopic
assessments.  
However, these results are in line with the low \ciso\ values obtained
for other globular clusters such as 
M3 (\ciso \ $\approx$ 6, \citealt{pilachowski_carbon_2003}), 
M4 (\ciso \ $\approx$ 5, \citealt{brown_high-resolution_1989}), and 
M22 (\ciso \ $\approx$ 6, \citealt{smith_carbon_1989}).
Assuming a \ciso\ value of 6, we then were able to estimate [C/Fe] $\sim$ -0.5, 
and [N/Fe] $\sim$ 1.
Figure \ref{cn_comp} shows the co-added RGB tip spectra and a corresponding 
synthesis for the CN region around 3880 \AA. Noise limitations are evident, 
even in these co-added spectra. 
\ion{C}{1} and \ion{N}{1} abundances will not be further considered in this paper 
given the difficulties in trying to determine these abundances.

\textit{Oxygen and Sodium: }
The \ion{O}{1} abundances for the RGB stars in our sample were determined
by synthesis of the [\ion{O}{1}] 6300.3 \AA\ line, while for three of our 
RHB and all of the BHB stars, oxygen abundances were determined from the 
\ion{O}{1} triplet at 7771.94 \AA, 7774.17 \AA, and 
7775.39 \AA . 
NLTE corrections are negligible for any forbidden transitions, but 
such effects can be quite large for abundances derived from the \ion{O}{1} 
triplet (eg., \citealt{sitnova_influence_2013} and references therein).
We used the estimates given in their Table 11 to extrapolate NLTE 
corrections for our final temperatures by deriving a logarithmic relation 
between these corrections and temperature, which is given by: 
\[ \mathrm{[O/Fe]_{NLTE}-[O/Fe]_{LTE}} = -1.43 \ \mathrm{log_{e}}(\teff) + 12.066. \]
NLTE corrections for our stars range from $-$0.35 dex to $-$0.76 dex.

\ion{Na}{1} abundances in our stars were derived using 
\textit{EW} measurements from four \ion{Na}{1} lines: 5889.95 \AA, 
5895.92 \AA \ (the D lines), 8183.26 \AA, and 8194.82 \AA.
The \textit{EWs} for these lines can be found in Table \ref{line_list}. 
We derived NLTE corrections using the INSPECT\footnote{
www.inspect-stars.net} 
web interface, which is based on \cite{lind_non-lte_2011}. 
Unfortunately, the parameters of some of our stars exceed the \teff - \logg \ 
limits. For such parameters, we extrapolated the NLTE corrections
given by the websites to our specific \logg \ and \teff\ combinations.
In the case of RGB stars, the limit was set by \logg \ as the website did 
not accept values lower than 1.00. For the BHB stars, the limit was set by \teff.
NLTE corrections obtained is this manner should not be viewed 
as definite, but rather as zeroth order estimates.

From Figure~\ref{light_elements_1}, it seems that \ion{O}{1} abundances are enhanced in our 
RGB stars as compared to other evolutionary stages and previous studies. 
However, since we are using different features to determine the \ion{O}{1} 
abundances in different evolutionary groups, caution is warranted 
when interpreting this abundance trend. 
In Figure~\ref{o_na_carretta}, we compare \ion{Na}{1} and \ion{O}{1} abundances.
The anti-correlation of these two elements, which seems to be exclusive to
GCs, has been confirmed by many studies (Car09, \citealt{gratton_abundance_2004} and 
references therein). It is currently believed that 
this trend results from pollution of the cluster medium by an earlier generation of stars 
(\citealt{gratton_abundance_2004}).
Unfortunately, we cannot confidently say that we observe such an 
anti-correlation in our sample, since we were only able to derive 
both \ion{Na}{1} and \ion{O}{1} abundances for 11 out of our 25 stars.
However, the abundances derived for these stars
fall within the same general ranges, perhaps slightly elevated, 
as those of Car09 (grey symbols), which consist of some of the most
metal-poor GCs known: NGC 6397 ([Fe/H] $\approx$ -2.02), 
M55 ([Fe/H] $\approx$ -1.94), M15 ([Fe/H] $\approx$ -2.37), 
and M30 ([Fe/H] $\approx$ -2.27). 
All quoted metallicities were obtained from \cite{harris_catalog_1996}.
Data for stars 117 and 160 (see \S\ref{finalparams} for
more details) are shown in green again. 
These two stars seem to have particularly high [O/Fe] values, which, in 
light of the problems associated with deriving their atmospheric parameters, 
are no cause for serious concern.
We conclude that, on average, M68 is not overabundant 
in sodium, but our data does suggest an oxygen overabundance.
Additionally, if our data is combined with the Car09, we clearly see an anti-correlation
of our derived oxygen and sodium abundances.
A closer look at Figure~\ref{o_na_carretta} also
appears to reveal a systematic difference of \ion{O}{1} abundances
between the RGB, RHB, and BHB. Such differences have
been observed and analyzed by several other authors (see
\citealt{marino_sodium-oxygen_2011} and references therein). 
Given that we employed different lines to obtain
our abundances of each of the evolutionary stages and relied on crude 
corrections to account for NLTE effects, 
we will not attempt to dissect the details of these differences
here.

We now compare our oxygen abundance results with those of Lee05 and Car09.
Car09 derived similar oxygen abundances also using the 
6300.3 \AA\ [\ion{O}{1}] transition for several different GCs.
The average metallicity for the Car09 M68 stars is slightly
higher than ours, and since they adopted purely photometric \logg \ 
and \teff \ values, their slightly lower \ion{O}{1} abundances are expected. 
In the specific case of the two overlapping RGBs,
the differences are $\Delta$[O/Fe]$_{57}$\footnote{
$\Delta$[O/Fe] = [O/Fe]$_{\mathrm{previous \ study}}$ $-$ [O/Fe]$_{\mathrm{this \ study}}$}
= $-$0.35 for star 57, while for star 79 we could not derive an \ion{O}{1} abundance.
The discrepancy between our results and those of Lee05 are more 
difficult to understand. For star 160, $\Delta$[O/Fe]$_{160}$ = $-$0.43 and for star 117 
$\Delta$[O/Fe$_{117}$] = $-$0.53. 
However, Lee05 employed a very different atmospheric parameter derivation 
methodology. Furthermore, they did not force ionization balance between 
neutral and ionized Fe lines, resulting in \ion{Fe}{2} 
abundances that are on average, 0.37 dex higher than the corresponding
\ion{Fe}{1} abundances. They chose to normalize their oxygen abundances
using these elevated \ion{Fe}{2} values, naturally leading to higher
[O/Fe] ratios.

The differences between our sodium abundances and those of Car09 are 
$\Delta$[Na/Fe]$_{57}$ = $+$0.17, and $\Delta$[Na/Fe]$_{79}$ = $+$0.14 . 
These differences can mostly be attributed to the slightly different
parameter derivation methodology and NLTE correction algorithm 
employed by Car09. 
The discrepancies between our work and that of Lee05 are a bit greater: 
$\Delta$[Na/Fe]$_{117}$ = $-$0.27 and $\Delta$[Na/Fe]$_{160}$ = $+$0.36.
Again, these differences, considering the disparate approaches, are negligible. 
We note here that \ion {Na}{1} NLTE corrections could not be derived for
star 160 using the methods described above. 

\textit{Aluminum: } 
All of our aluminum abundances were derived from the \ion{Al}{1} resonance lines at 3944.01~\AA \ 
and 3961.53~\AA. Due to severe NLTE effects, these lines are 
known to give abundances which are, on average, a factor of 6 lower than those 
derived from other \ion{Al}{1} lines (e.g., 
{\citealt{baumueller_aluminium_1997},
\citealt{andrievsky_nlte_2008}).
Unfortunately, these are the only measurable \ion{Al}{1} lines in our spectra.
In their Figure 2, \citealt{andrievsky_nlte_2008} give NLTE corrections 
applicable to some of our giants.
We estimate corrections of 0.4 dex for stars 79 and 172, a correction of
0.55 dex for stars 226 and 256, and a correction of 0.5 for star 440. 
Atmospheric parameters for all other stars in our sample are outside
the recommended limits of \cite{andrievsky_nlte_2008} and therefore are
not considered for NLTE corrections. Table~\ref{al_nlte} gives the NLTE
\ion{Al}{1} abundances, as well as NLTE corrected \ion{Na}{1} abundances.
Clearly, there is a spread amongst \ion{Al}{1} abundances in M68. 
However, all low/high [Al/Fe] ratios correspond to low/high [Na/Fe] ratios.
Figure \ref{al_na} shows this well-known correlation between [Al/Fe] and [Na/Fe] for globular clusters
(e.g Car09, \citealt{ivans_new_2001}, and \citealt{shetrone_al_1996}).
In the interest of having the maximum number of data points available, 
we decided to use the \ion{Al}{1} LTE abundances. Since
the NLTE corrections are almost constant for our stars (see discussion
above and Table~\ref{al_nlte}), there is no danger in destroying any
abundance correlations by using the LTE abundances.
The clearly positive correlation between these two elements provides 
evidence for primordial abundance enhancements by earlier
generations of stars. We will not attempt 
to explore these primordial abundance enhancements using our data,
but refer the reader to \cite{gratton_abundance_2004} for more information.

Lee05, who derive $\langle\mathrm{[Al/Fe]}\rangle$ = 1.08, and Car09, 
who derive $\langle\mathrm{[Al/Fe]}\rangle <$ 0.74, measured their abundances
using the subordinate \ion{Al}{1} lines at 6696.03 \AA\ and 6698.67 \AA.
We note that Car09 explicitly state that all their \ion{Al}{1} abundances
are upper limits, which, in combination with the statement 
by \cite{andrievsky_nlte_2008} that
these subordinate lines should not be visible at [Fe/H] $\lesssim$ $-$2.5,
leads us to lend less significance to the differences between our study and
Lee05. We derive 
$\langle\mathrm{[Al/Fe]_{NLTE}}\rangle$ = 0.31, which
given the difficulties explained above, seems in concord with Car09.
To ensure that our lack of detection of these subordinate lines is not caused by noise limitations, 
we co-added RGB spectra of the stars with the highest \ion{Al}{1} abundances
implied by the resonance lines. Despite these efforts, we were not able to measure the subordinate lines,
or \ion{Al}{1} line at 7836.14 \AA, which suggests that [Al/Fe] $<$ 0.78. 

\textit{$\alpha$ Elements: }
The $\alpha$ and $\alpha$-like elements considered in this study 
are \ion{Mg}{1}, \ion{Si}{1}, \ion{Ca}{1}, and \ion{Ti}{1}/\ion{Ti}{2}. 
Their abundances, which were obtained from $EW$ measurements, 
are compared in Figure~\ref{light_elements_1}.
We will discuss them in order of increasing $Z$.

\ion{Mg}{1} abundances for all evolutionary stages show good
agreement, which, in combination with Figure~\ref{mg_al}, where we see 
the expected [Mg/Fe]$-$[Al/Fe] anti-correlation, 
supplies further evidence for primordial abundance
variations in GC stars. Dissecting the details of these abundance behaviors
is beyond the scope of this paper. Here, we mainly demonstrate that our atmospheric parameter
derivation approach reproduces all the classical abundance hallmarks of any GC study.
Star 202, with [\ion{Al}{1}/Fe] $\sim$ 0.4 in Figure~\ref{mg_al} 
is the hottest RHB in our sample, and with \teff\ = 6257, very close
to the RR-Lyrae gap. Therefore, the \ion{Al}{1} abundances 
for this star could be compromised. We have also chosen not
to include BHB Mg abundances in Figure~\ref{mg_al} due to 
reasons listed in \S\ref{bhb}.

Car09, Lee05 and \cite{behr_chemical_2003} also derived \ion{Mg}{1} abundances for
their stars. They differ from our measurements
as follows: $\Delta_{Mg / Fe, 57}$ = -0.044,
$\Delta_{Mg / Fe, 79}$ = -0.083,  
$\Delta_{Mg / Fe, 117}$ = -0.25,
and $\Delta_{Mg / Fe, 160}$ = -0.01.
All of these ratios show excellent agreement, except those of star 117. 
The large discrepancy for this star, however, is no real surprise given the hugely different approaches 
for determining atmospheric parameters and the fact that this particular
star might be an example of an extreme RGB tip star, making its analysis very difficult.
We will not compare the individual stellar abundances for star 324
with those of \cite{behr_chemical_2003}, since we were unable to
derive proper atmospheric abundances for this particular star. 
\cite{behr_chemical_2003} derive a $\langle$[Mg/Fe]$\rangle$ = 0.18,
while we obtain $\langle$[Mg/Fe]$\rangle$ = 0.35 for our BHBs. 
This difference can be explained by the different approach
taken by \cite{behr_chemical_2003} to derive their atmospheric parameters.

Unlike the [Mg/Fe] ratios, our derived [Si/Fe] abundances show
large discrepancies between the evolutionary stages. This is because,
as several studies have shown, \ion{Si}{1} abundances
drop with increasing \teff \ (e.g. Figure 10 of 
\citealt{preston_atmospheres_2006} and references therein). 
In the interest of consistency, we derived all of our \ion{Si}{1}
abundances from the 3905.53 \AA \ line, which is saturated in 
the RGB, leading to possibly unreliable abundances. If we had
used the 4102.94 \AA \ \ion{Si}{1} line for this evolutionary stage, 
our RGB abundances would be elevated by about 0.4 dex. 
Unfortunately, the 4102.94 \AA \ line is not measurable in the 
RHBs or BHBs.

For the RGB branch, our study yields $\langle$[\ion{Si}{1}/Fe]$_\mathrm{RGB}\rangle$ = 
0.40 (see Table~\ref{evolution_abund}),
which, if compared to $\langle$[\ion{Si}{1}/Fe]$_\mathrm{RGB}\rangle$ values from 
other metal poor clusters:  
$\langle$[\ion{Si}{1}/Fe]$_\mathrm{RGB, M15}\rangle$ = 0.55, and 
$\langle$[\ion{Si}{1}/Fe]$_\mathrm{RGB, M92}\rangle$ = 0.59
(both from \citealt{sneden_barium_2000}), seems slightly too low. 
However, \citealt{sneden_barium_2000} use only the 5948.55 \AA \ \ion{Si}{1} 
line, and a different set of $gf$ values. Similar to \cite{sneden_barium_2000}, 
Lee05 also derive an analogously high \ion{Si}{1} abundance, 
$\langle$[\ion{Si}{1}/Fe]$_\mathrm{RGB, M68}\rangle$= 0.71,
which they base on \textit{EW} measurements of 6-7 m\AA \ for all their stars.
Given that their \textit{S/N} is comparable to ours, noise limitations could 
contribute to their derived Si overabundances.
To ensure that our \textit{EW} measurements
are not at fault, we calculated the \textit{EW} values needed to reproduce 
the Lee05 abundance for several different high excitation \ion{Si}{1}
lines, including those used by Lee05. 
Subsequently, we inspected the spectra of our stars for these lines
and tried to measure the \textit{EW} needed to reproduce the Lee05 abundances. 
We were unable to measure or visually locate
any of these lines in our spectra, confirming that M68 most likely
does not exhibit an Si overabundance. In their study, 
Car09 derive $\langle$[Si/Fe]$\rangle$ = 0.40, in good agreement with our derived value.

Our RHB Si measurements yield $\langle$[Si/Fe]$\rangle$ = 0.16, 
which is in accord with \cite{preston_atmospheres_2006}, 
who derive $\langle$[\ion{Si}{1}/Fe]$_\mathrm{RHB, M15}\rangle$ = 0.32.
While we expect lower abundances with increasing \teff\ (see above),
a drop of 0.5 dex in [Si/Fe] between our average RHB abundances and
those of Lee05 would be outside of the usual range found by previous studies.

As a final remark on \ion{Si}{1} abundances, we perform
a star-by-star comparison with Car09, which yields
these differences: 
$\Delta_{Si / Fe, 57}$ = $-$0.23,
$\Delta_{Si / Fe, 79}$ = $-$0.051,
The reasons for the large differences are explained above and can 
also be attributed to the different methodologies employed for deriving
these abundances and associated atmospheric parameters.
We were unable to derive Si abundances for either stars 117 or
160, precluding us from a comparison with Lee05.

Perhaps a more important and reliable diagnostic than Si
is Ca, which exhibits more measurable transitions and has been shown
to give constant abundances for GCs with [Fe/H] $<$ -1.00. 
Our analysis yields $\langle$[Ca/Fe]$_{\mathrm{M68}}\rangle$
= 0.36, very much in line with previous studies shown
in Figure 4 of \cite{gratton_abundance_2004}.
They derive $\langle$[Ca/Fe]$\rangle$
=+0.25 ($\sigma$~=~0.02) for 28 clusters with [Fe/H] $<$ -1.00.
A comparison with Lee05, who also derive 
\ion{Ca}{1} abundances yields the following: 
$\Delta_{Ca / Fe, 117}$ = -0.16, 
and $\Delta_{Ca / Fe, 160}$ = -0.04.}
For a thorough 
review on Ca abundances in GCs,
please see \cite{gratton_abundance_2004}.

As a final $\alpha$ element, we measured the abundances of \ion{Ti}{1}
and \ion{Ti}{2}, whose
abundances provide a good diagnostic of possible over-ionization in stellar 
atmospheres.
We will not compare our Ti abundances to those of Lee05, since the
methodology of the two studies are too disparate.
A quick look at Figure~\ref{light_elements_1} reveals over-ionization in RGB stars
(\ion{Ti}{2} abundances are, on average, 0.34 dex higher), 
while the agreement between Ti I and Ti II is excellent for RHB stars.
For reasons explained above, there are no measurable
Ti I lines in our BHB stars.
Overall, our data shows an $\alpha$ element enhancement, 
as expected from globular clusters. A similar enhancement in M68
was found by \cite{behr_chemical_2003}.

\textit{Fe-peak Elements: }
\ion{Sc}{2}, \ion{Cr}{1}, \ion{Cr}{2}, 
\ion{Mn}{1}, \ion{Mn}{2}, and \ion{Co}{1} make up the $Fe$-peak elements
measured in this study.
Their combined mean abundances are 
$\langle$[X/Fe]$_{Fe \ \mathrm{peak}}\rangle$ = 0.11 ($\sigma$ = 0.30). 
Figures~\ref{light_elements_1}~and~\ref{light_elements_2} 
show the behavior for these elements amongst different evolutionary stages.
\ion{Sc}{2} abundances agree for all our evolutionary states
and the Lee05 data. The large difference between
\ion{Cr}{1} and \ion{Cr}{2} abundances could be a cause for concern, since 
we required ionization equilibrium for our abundances. However, 
such behavior has been observed in other cluster as well as 
field stars and is
currently unexplained (see \citealt{preston_atmospheres_2006} and
references therein). In Figure~\ref{light_elements_2}, we show 
a comparison with the M92 data of \cite{langer_spectroscopic_1998}. 
Clearly, their average abundance is much lower than ours. Due to 
the known problems with this element, however, we will not lend too much
weight to this difference.

\ion{Mn}{1} and \ion{Mn}{2} abundances also seem to show 
over-ionization and large variations amongst individual 
stars. However, nearly all of these abundances are based on a single
stellar line measurement. We therefore urge the reader to treat these
abundances with caution.
\ion{Co}{1} abundances show agreement between the RGB and RHB branches, but are again
much lower than those of \cite{langer_spectroscopic_1998}. The reasons 
for this behavior are unknown.

Lee05 measured \ion{Mn}{1} and \ion{Sc}{2}, and for these 
species our differences are:
$\Delta_{Sc / Fe, 117}$ = -0.24,
$\Delta_{Sc / Fe, 160}$ = -0.26,
$\Delta_{Mn / Fe, 117}$ = 0.03, and  
$\Delta_{Mn / Fe, 160}$ = -0.04.
Unfortunately, no other abundance comparisons are possible.

\textit{Copper and Zinc: }
The last two elements in Figure~\ref{light_elements_2} are \ion{Cu}{1} and
\ion{Zn}{1}. Lee05 also derived \ion{Cu}{1} abundances and the difference between
the overlapping star 160 is $\Delta_{Cu / Fe, 160}$ = -0.55. These differences can be
attributed to the disparate approaches in deriving atmospheric parameters.
For \ion{Zn}{1}, no previous references exist.

\subsection{Heavy Elements\label{heavy_ele}}
\textit{n-capture Elements: } Figure~\ref{heavy_elements} shows box plots 
of the distribution of \textit{n}-capture elements in our stars. 
Unlike many previous studies, 
the violet extent of our spectral coverage
allows us to include rare-earth-element
species such as \ion{Dy}{2} and \ion{Yb}{2}. The distribution
of \textit{n}-capture elements can be used to infer $r$- or
$s$-process enrichment (see Figures 1, 2, and 10 in 
\citealt{sneden_neutron-capture_2008}). As part of
$r$-process enrichment, we should observe 
low [Ba/Eu] and [La/Eu] ratios, both of which are clearly
present in our sample. We therefore conclude that
M68, much like other metal poor GCs 
(e.g. \citealt{sobeck_abundances_2011}), 
exhibits $r$-process enrichment. The constant 
offset of $\sim$ 0.26 dex between the RGB and RHB
evolutionary stages can be attributed to the usage of different {\sf MOOG} versions
(see \S\ref{finalparams} for more details),
as well as general difficulties associated
with the size of the parameter space covered by this study. 

Figure~\ref{heavy_elements} also demonstrates discrepancies
between \ion{Sr}{2} and \ion{Ba}{2} abundances amongst the evolutionary stages.
These are easily explained. For \ion{Sr}{2}, the employed lines at 4077.71 \AA\ and 4215.52 \AA\
are saturated in almost all RGB and RHB stars, resulting in unreliable abundances.
In the hottest RHB and all BHB stars, these \ion{Sr}{2} lines are unsaturated and
yield consistent abundances. The \ion{Ba}{2} discrepancies can be traced back to the usage of different
sets of lines for obtaining abundances. The well known 4554.03 \AA \ \ion{Ba}{2} line
is saturated in our RGB and RHB stars, and we used the 
5853.69 \AA, 6141.73 \AA, 6496.91 \AA\ in these stars. 
In our BHB stars, these \ion{Ba}{2} lines were 
not available, so we reverted to
the 4554.03 \AA\ line. 

Car09 did not derive any 
\textit{n}-capture process abundances, and therefore we will compare all of our 
abundances to those of Lee05.
The differences for star 117 are 
$\Delta_{Ba \ II/Fe, 117}$ = -0.28 (only element available 
for comparison),
while those for star 160 are 
$\Delta_{Ba \ II/Fe, 160}$ = -0.29, 
$\Delta_{La \ II/Fe, 160}$ = -0.25, and 
$\Delta_{Eu \ II/Fe, 160}$ = -0.14. Like before, 
$\Delta_{X/Fe}$ is defined to be [X/Fe]$_{previous \ study}$ $-$ [X/Fe]$_{this \ study}$.
Possible reasons for the differences between our abundances and those
derived by Lee05 have been discussed in \S\ref{light_ele}.

\subsection{Abundance Uncertainties\label{uncert}}
Much like atmospheric parameter uncertainties, we
derived representative abundance $\sigma$ values for each  
evolutionary state in our sample. The representative
stars remain the same: 472 (RGB tip), 172 (lower RGB), 
36 (RHB), and 337 (BHB). We note that we derived 
\ion{Cr}{1} abundance uncertainties for the BHBs using star 
289, since we could not measure any \ion{Cr}{1} lines in 
star 337.

To obtain uncertainties for elements whose abundances
are based on \textit{EW} measurements, 
we took the following approach:
Uncertainties for any element for which more than 3 lines were measured, the 
standard deviation of the abundances implied by the measured lines was used. 
For any element with less than 3 lines, we re-measured the \textit{EW} 
now considering factors such as continuum placement and smoothing. 
Using these re-measured \textit{EW} values, we derived new abundances, which
allowed us to calculate uncertainties for the corresponding species.
This method led to uncertainties of $\simeq$ 0.13 dex for most elements.

For uncertainties for elements where spectral synthesis was used to obtain abundances, we re-synthesized
the spectra and again considered factors such as continuum placement, smoothing and assumed abundance.
Just like with our \textit{EW} uncertainties, this method leads to final $\sigma$ values of $\simeq$ 0.13 dex. 
The final uncertainties obtained in the above 
described fashions can be found in Table \ref{abund_uncert}. It is also mentioned here 
that our uncertainties in \teff, \logg, and \vmicro \ lead to additional uncertainties 
of 0.2 - 0.3 dex, which are not included in the 
results of Table \ref{abund_uncert}. 

\section{CONCLUSIONS\label{conclusions}}
In this paper, we have explored the atmospheric parameters
and detailed chemical compositions of 25 evolved members of M68.
Particular attention has been paid to comparisons of the assets and 
liabilities of photometrically based and spectroscopically based parameters.  
From the discussion in \S\ref{finalparams}, and the evidence presented
in \S\ref{comparison_photo_spect}, it is now clear that 
difficulties in deriving atmospheric parameters in an internally consistent 
manner over a parameter space that covers $\sim$ 4000 K in \teff, and 
$\sim$ 3.00 dex in \logg\ exist and need to be accounted for.
Luckily, for M68, as for almost any cluster, we have reliable reddening 
and distance moduli that allowed us to treat these problems by creating 
a hybrid spectroscopy-photometry approach.

\begin{figure}
\epsscale{0.8}
\plotone{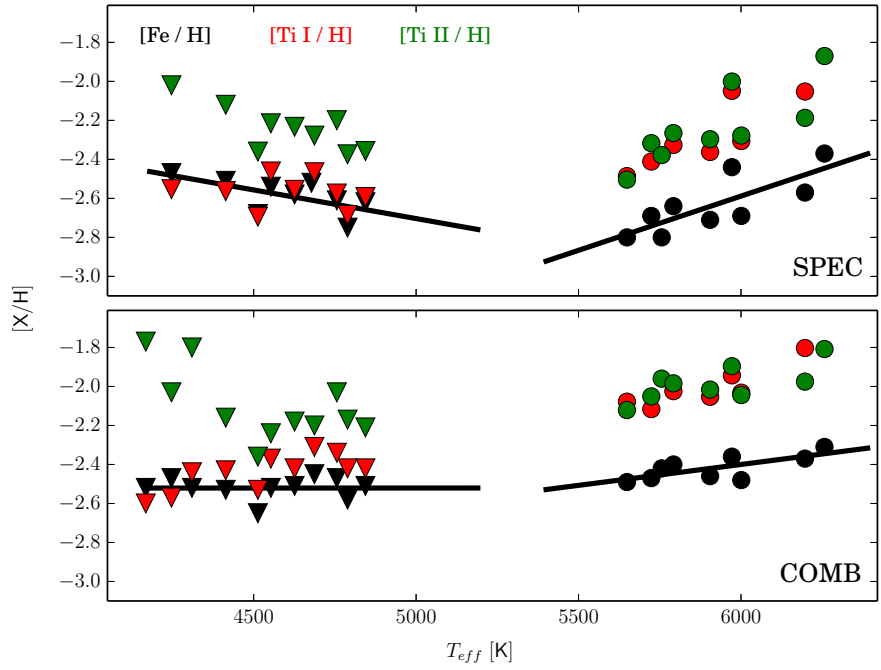}
\caption{
   \footnotesize{Comparison of ionization and metallicity effects using different atmospheric parameters.
   The solid black line represents a linear fit to [Fe/H] of RGB and RHB stars as a function of temperature.}
}
\label{ionization_comp}
\end{figure}

\begin{figure}
\epsscale{0.8}
\plotone{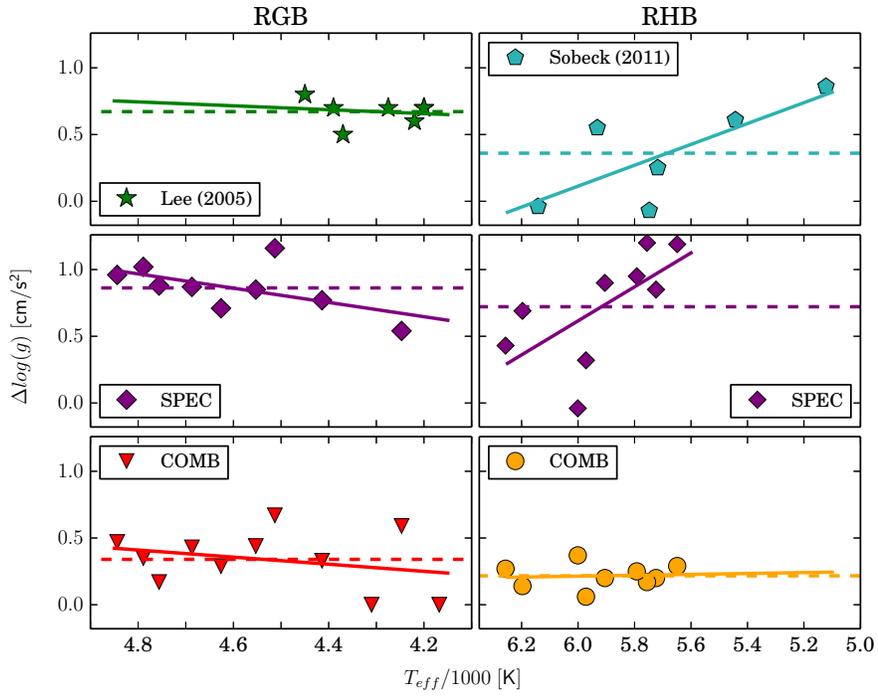}
\caption{
  \footnotesize{Comparisons of $\Delta\logg \ (\equiv \logg_{PHOT} - \logg_{comparison \ value})$ values as
   a function of temperature for different studies and methodologies. The solid lines represent a linear fit to the data, while dashed lines
   give the average $\Delta\logg$ offset.}
}
\label{logg_comp_rgb_rhb}
\end{figure}

\begin{figure}
\epsscale{0.8}
\plotone{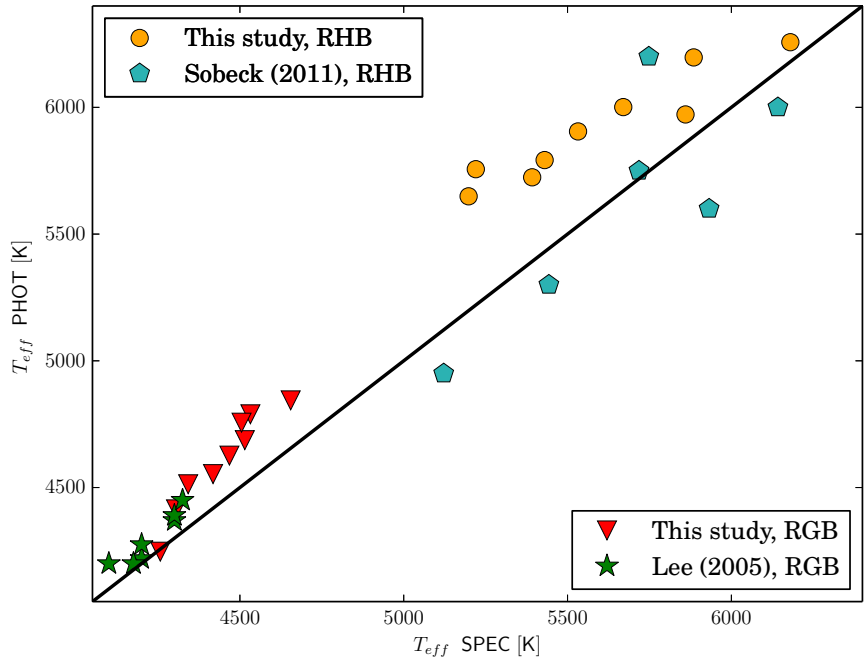}
\caption{
   \footnotesize{Comparison of photometric and spectroscopic temperatures. 
   Our data agrees well with Lee05, but differs quite significantly from Sob11.}
}
\label{teff_comp_rgb_rhb}
\end{figure}

\begin{figure}
\epsscale{0.8}
\plotone{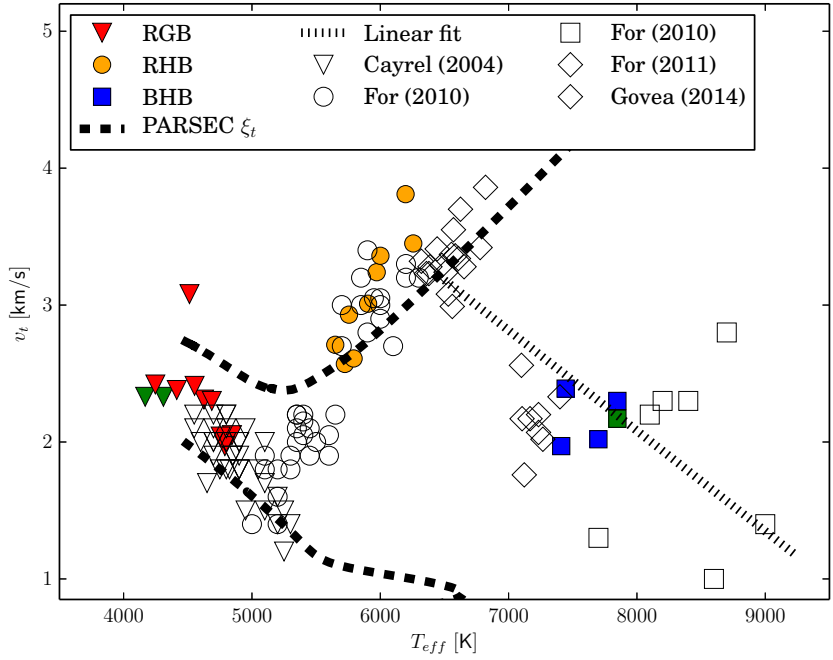}
\caption{
   \footnotesize{Microturbulent velocities as a function of evolutionary stage. Our results agree well with previous studies
   and the \vmicro \ calibrations of \cite{gratton_abundances_1996}. In this Figure, we have used a diamond to represent 
   the RR-Lyrae evolutionary stage. Solid symbols represent this study, while
   unfilled symbols represent comparison values from previous studies. See 
   \S\ref{comparison_photo_spect} for more details.}
}
\label{vt_comparison}
\end{figure}

\begin{figure}
\epsscale{0.8}
\plotone{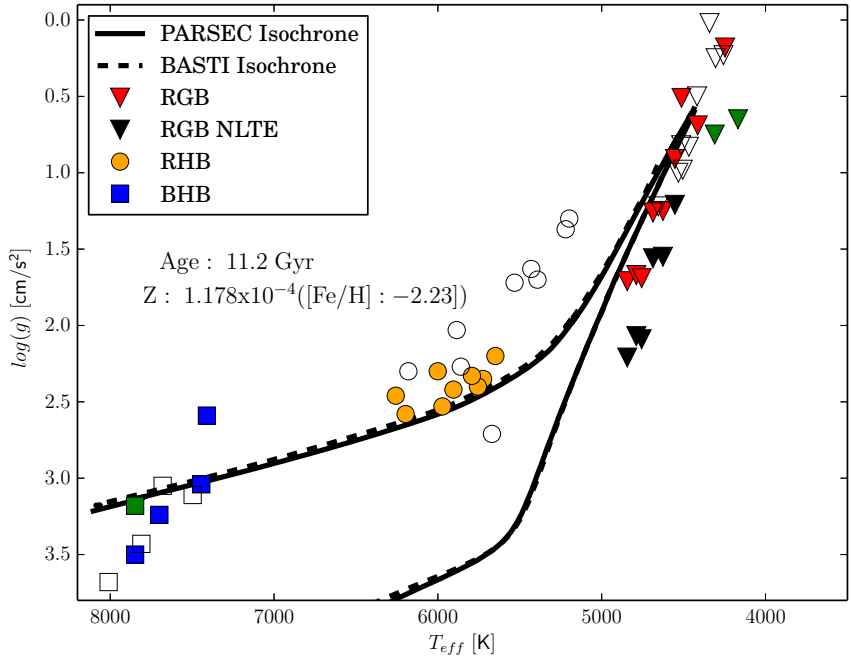}
\caption{
   \footnotesize{Comparison of isochrone tracks with derived atmospheric parameters. Green markers
    indicate derivation of atmospheric parameters by photometric data only. Markers without fill indicate 
    purely spectroscopic parameters. See \S\ref{finalparams} for more information.}
}
\label{isochrone}
\end{figure}

\begin{figure}
\epsscale{0.8}
\plotone{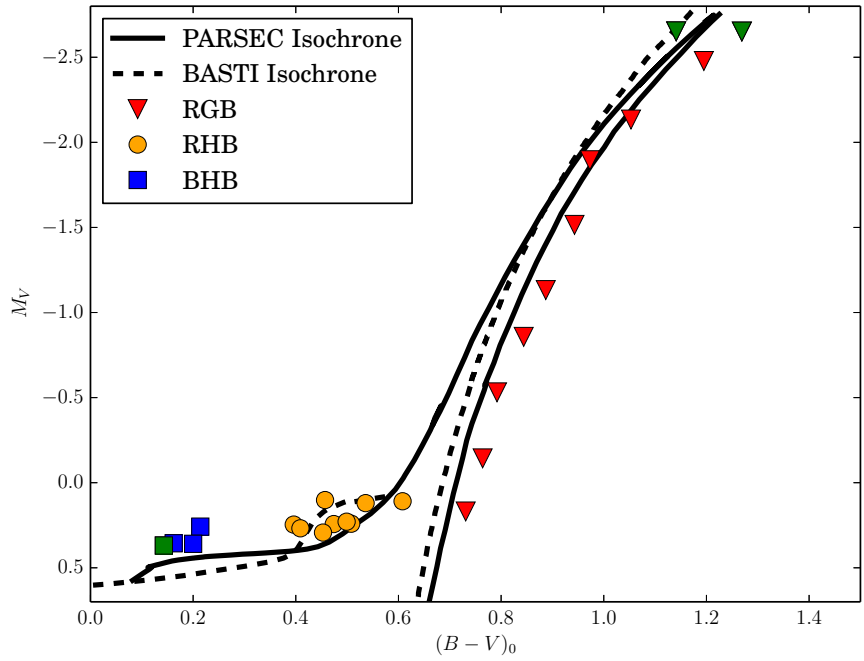}
\caption{
   \footnotesize{Reproduction of Figure \ref{isochrone}, but in color space. The fit of the PARSEC isochrones
   is somewhat better than that of the BaSTI tracks.}
}
\label{isochrone_color}
\end{figure}

Some of the discovered weaknesses of the two standard approaches
include photometric derivation of \logg\ values.
As mentioned in Sec. \ref{initial}, the physical \logg \ 
depends on the mass of the star in consideration. 
Unfortunately, precise mass loss between the turn-off, RGB and later 
evolutionary stages is still uncertain. Our method of applying a single mass to all stars in our sample provides a 
good zeroth order approximation of the physical \logg, 
but for increasingly detailed studies, more accurate stellar masses should be adopted.
Spectroscopic weaknesses include scarcity of lines
in hotter RHB and BHB stars, 
as well as serious metallicity trends (see Figure~\ref{ionization_comp}).
A combination of both the photometric and spectroscopic methods
allowed us to address most of these problems. One of the main highlights
of our final adopted approach is the lack of any obvious metallicity difference
between evolutionary stages (see Figure~\ref{ionization_comp} or
$\langle$[Fe/H]$\rangle$ values for RGBs and RHBs in Table~\ref{evolution_abund}).
Our analysis results in $\sigma_{\mathrm{\langle[Fe/H]\rangle}}$ = 0.14, for
all stars, including the BHBs. We remind the reader that this lack of metallicity difference is not
inherent to our analysis or our atmospheric parameter derivation
method. Instead, it naturally grows out of our adoption of
photometric temperatures, making this result even more remarkable.

In addition to the metallicities being in good agreement, Table~\ref{evolution_abund}
shows that the abundances for all other elements agree across the different
evolutionary stages. For the elements that seem to exhibit any discrepancies,  
a valid and detailed explanation is given in \S\ref{abunds}. 
We can therefore conclude that even though
we adopted a non-standard, hybrid approach to deriving our atmospheric parameters, the resulting
abundances are what one would expect from a classical purely spectroscopic analysis.
Moreover, we were also able to reproduce all classical hallmarks of GC populations, such 
as the [Na/Fe]$-$[O/Fe] and [Mg/Fe]$-$[Al/Fe] anti-correlations,
Ca and Si abundances that agree with previous studies, as well 
as slight $r$-process enrichment.
In the absence of very detailed NLTE calculations and/or 3D model atmospheres, such a hybrid 
approach may be necessary for us to further develop our understanding of cluster stars,
at least from a stellar atmospheric perspective.

\nocite{wujec_transition_1981}
\nocite{wiese_atomic_1969}
\nocite{wiese_new_2006}
\nocite{wiese_atomic_1996}
\nocite{wickliffe_atomic_2000}
\nocite{malcheva_radiative_2006}
\nocite{lawler_improved_2001}
\nocite{lawler_improved_2008}
\nocite{lawler_experimental_2001}
\nocite{kelleher_atomic_2008}
\nocite{gallagher_oscillator_1967-1}
\nocite{caffau_photospheric_2008}
\nocite{biemont_lifetime_2011}
\nocite{den_hartog_improved_2003}

\acknowledgments
We thank Sloan Simmons for initial work on this project, 
as well as Karin Lind and Jeremy King for helpful discussions.
We also thank our referee for providing valuable suggestions and
thereby improving the quality of this paper.
Financial support for this research from the US National Science Foundation 
(grant AST-1211585) and the Rex G. Baker, Jr. Centennial Endowment to
the University of Texas are gratefully acknowledged.

\clearpage
\bibliographystyle{aa} % style aa.bst. 
\bibliography{m68_2014}
%\bibliography{chrisbib}

%%%%%FIGURES%%%%%%%
\clearpage

\clearpage

\clearpage

\clearpage

\clearpage

\clearpage

\clearpage

\clearpage

\clearpage

\clearpage
\begin{figure}
\epsscale{0.8}
\plotone{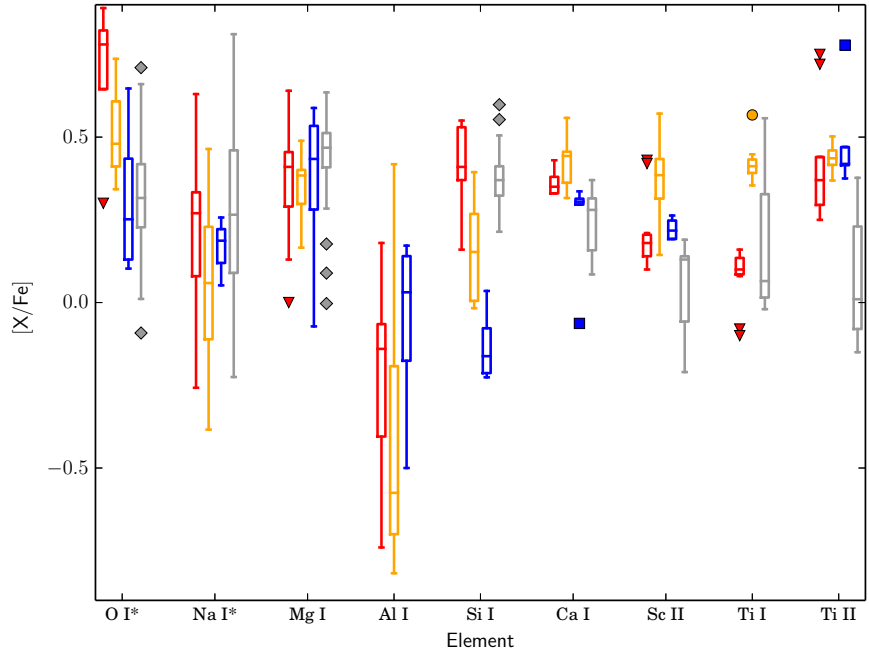}
\caption{
   \footnotesize{Comparisons of derived abundances for 8 $\leq Z \leq 22$. Any abundance differences
   between the evolutionary stages are explained in \S\ref{light_ele}. Solid markers outside the boxes represent
   outlier abundances. Grey boxes represent data from previous studies. Elements marked with `*' have NLTE
   corrections applied to them.
}
}
\label{light_elements_1}
\end{figure}

\clearpage
\begin{figure}
\epsscale{0.8}
\plotone{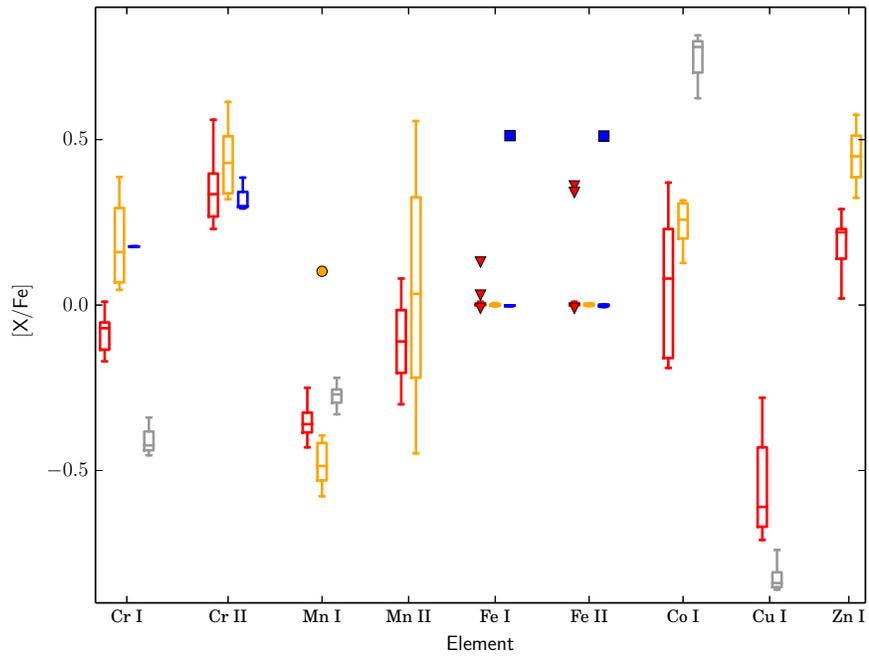}
\caption{
   \footnotesize{Continuation of Figure \ref{light_elements_1}, but with 24 $\leq \ Z \ \leq 30$. Abundance differences 
   are discussed in \S\ref{light_ele}.}
}
\label{light_elements_2}
\end{figure}

\clearpage
\begin{figure}
\epsscale{0.8}
\plotone{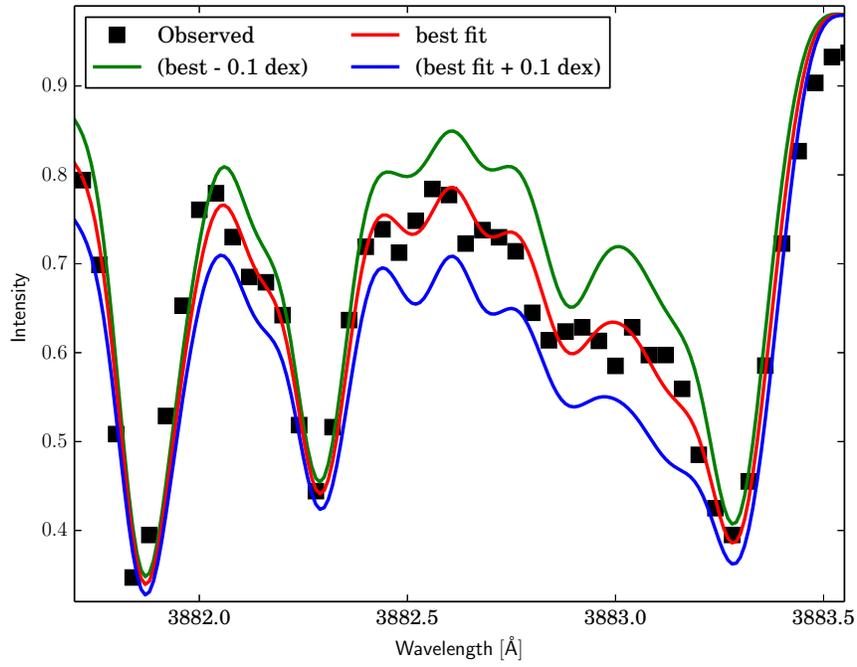}
\caption{
   \footnotesize{Synthesis of the CN region around 3880 \AA . Noise limitations are evident from this plot. The
   spectrum shown here is a co-addition of stars 117, 160, 472, and 481, all of which are RGB tip members.}
}
\label{cn_comp}
\end{figure}

\clearpage
\begin{figure}
\epsscale{0.8}
\plotone{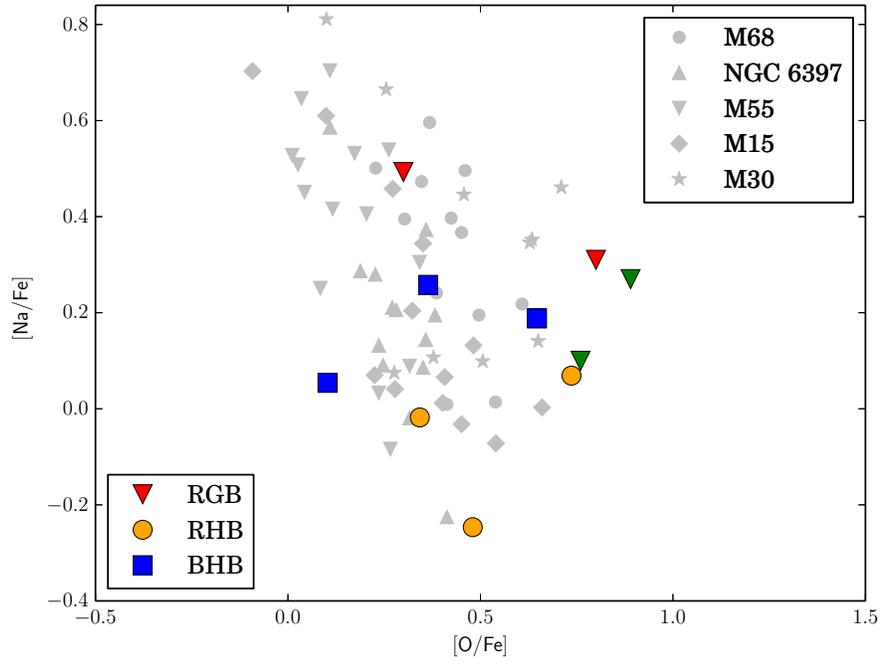}
\caption{
   \footnotesize{Correlation between [Na/Fe] and [O/Fe]. Our data is compared to that of Car09, who
   derived these abundances for RGBs in M68, NGC 6397, M55, M15, and M30. 
   NLTE corrections have been applied to for both O and Na abundances.}
}
\label{o_na_carretta}
\end{figure}

\clearpage
\begin{figure}
\epsscale{0.8}
\plotone{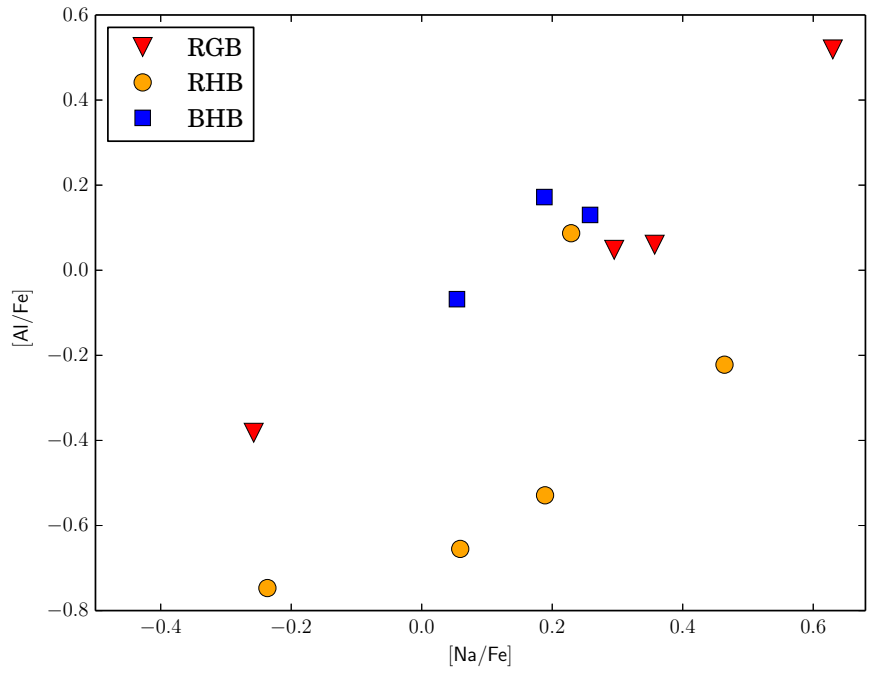}
\caption{
   \footnotesize{The classical comparison of [Na/Fe] to [Al/Fe], which produces the standard result for
   globular cluster studies. In this plot, NLTE corrections were applied only to Na.}
}
\label{al_na}
\end{figure}

\clearpage
\begin{figure}
\epsscale{0.8}
\plotone{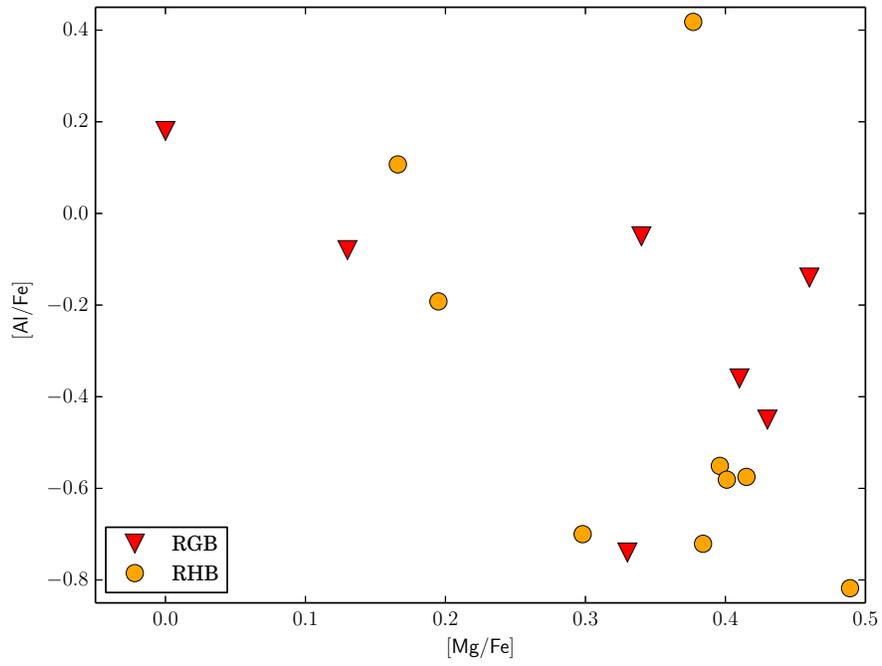}
\caption{
   \footnotesize{Anti-correlation of [Al/Fe] and [Mg/Fe] for stars in our sample.}
}
\label{mg_al}
\end{figure}

\clearpage
\begin{figure}
\epsscale{0.8}
\plotone{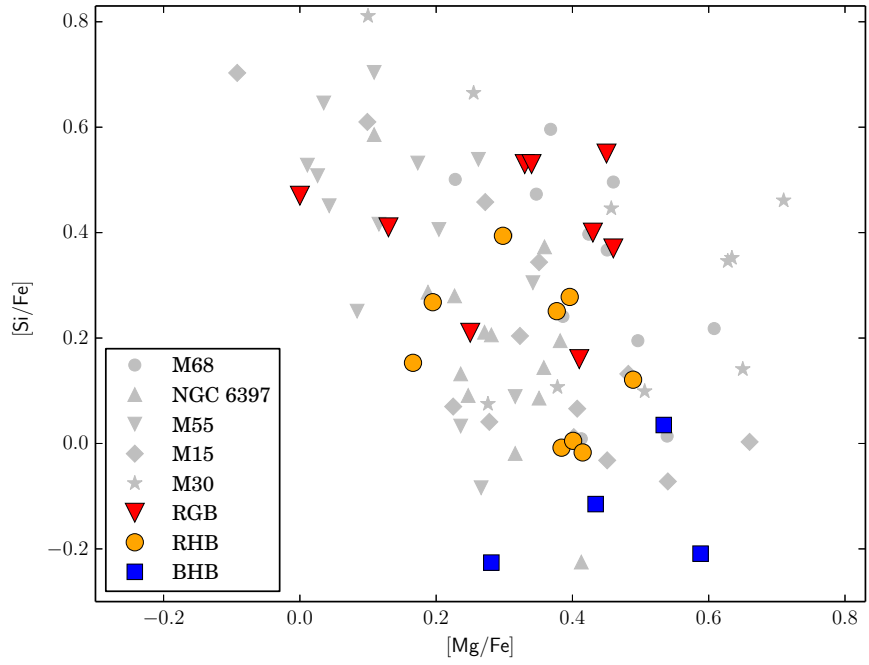}
\caption{
   \footnotesize{Correlation between Si and Mg abundances. The drop-off in the Si abundances with increasing
   temperatures is well known.}
}
\label{si_mg}
\end{figure}

\clearpage
\begin{figure}
\epsscale{0.8}
\plotone{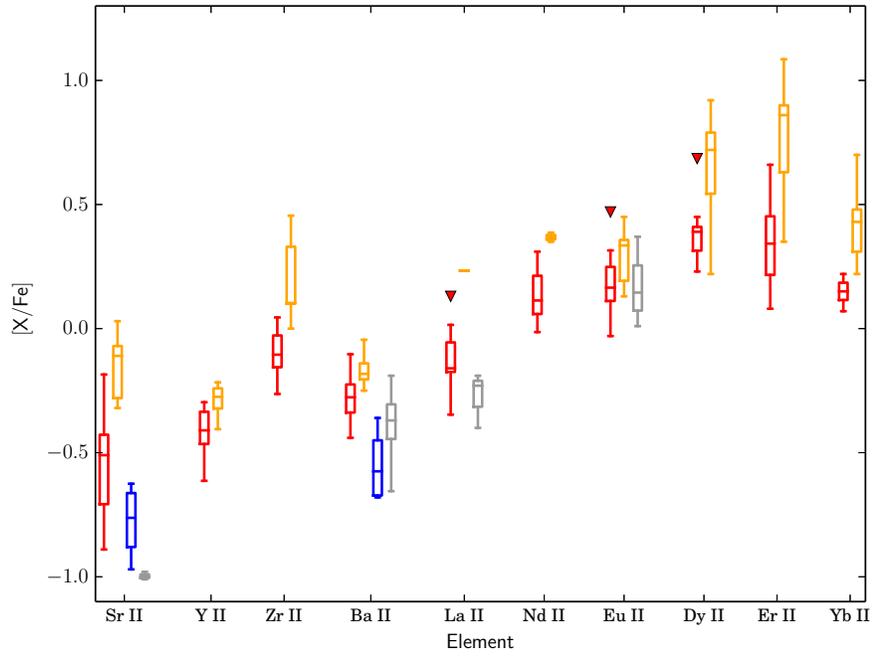}
\caption{
   \footnotesize{Abundances of \textit{n}-capture elements amongst the different evolutionary stages. A slight \textit{r}-process
   enrichment can be observed. Further details are given in \S\ref{heavy_ele}}
}
\label{heavy_elements}
\end{figure}

%%%%%%TABLES%%%%%%%
\clearpage
\begin{center}

\begin{deluxetable}{l r}
\tabletypesize{\footnotesize}
\tablecolumns{2}
\tablewidth{0pt}
\tablecaption{Fundamental parameters of NGC 4590\label{fun_params}}
\tablehead{
			\colhead {}												&
			\colhead{}													}

\startdata
Right ascension$^{1}$	& 		12 36 27.98 \\
Declination$^{1}$ 		& 		-22 44 38.6 \\
Distance$^{4}$			& 		10.3 kpc \\
Age$^{3}$					& 		11.2 Gyr \\
Metallicity$^{4}$	 		& 		-2.23 \\
$E(B-V)$$^{4}$	           	&      	     	0.05 \\
$(m-M)_{V}$$^{4}$	            &           	15.210 \\
\enddata

\tablenotetext{1}{\cite{goldsbury_acs_2010}}
\tablenotetext{2}{\cite{sollima_correlation_2008}}
\tablenotetext{3}{\cite{harris_catalog_1996}, (2010 edition)}

\end{deluxetable}
\end{center}

\clearpage
\begin{center}

\begin{deluxetable}{c c c c}
\tabletypesize{\footnotesize}
\tablecolumns{4}
\tablewidth{0pt}
\tablecaption{Photometric parameters\label{photometric_data}}
\tablehead{
			\colhead {Star}												&
			\colhead{\textit{V}}											& 
			\colhead{\textit{(B-V)}}										&
			\colhead{\textit{(V-I)}}										}

\startdata
\multicolumn{4}{c}{RGB photometry} \\					
\hline					
\textbf{57}	&	14.213	&	0.937	&	1.116\\
\textbf{79}	&	13.830	&	0.993	&	1.161\\
\textbf{117}	&	12.693	&	1.191	&	1.320\\
\textbf{160}	&	12.694	&	1.319	&	1.418\\
\textbf{172}	&	14.488	&	0.894	&	1.084\\
\textbf{226}	&	15.514	&	0.781	&	0.981\\
\textbf{256}	&	15.202	&	0.814	&	1.004\\
\textbf{440}	&	14.814	&	0.842	&	1.030\\
\textbf{450}	&	13.450	&	1.023	&	1.182\\
\textbf{472}	&	13.211	&	1.103	&	1.244\\
\textbf{481}	&	12.867	&	1.245	&	1.359\\ \\
\hline					
\multicolumn{4}{c}{RHB photometry} \\					
\hline					
\textbf{36}	&	15.588	&	0.524	&	0.723\\
\textbf{47}	&	15.545	&	0.567	&	0.775\\
\textbf{202}	&	15.591	&	0.446	&	0.676\\
\textbf{334}	&	15.586	&	0.558	&	0.727\\
\textbf{403}	&	15.465	&	0.586	&	0.798\\
\textbf{454}	&	15.447	&	0.507	&	0.740\\
\textbf{458}	&	15.638	&	0.503	&	0.700\\
\textbf{533}	&	15.574	&	0.549	&	0.746\\
\textbf{547}	&	15.613	&	0.459	&	0.685\\ \\
\hline					
\multicolumn{4}{c}{BHB photometry} \\					
\hline					
\textbf{170}	&	15.714	&	0.193	&	0.323\\
\textbf{289}	&	15.603	&	0.264	&	0.295\\
\textbf{324}	&	15.772	&	0.204	&	0.287\\
\textbf{377}	&	15.701	&	0.212	&	0.348\\
\textbf{391}	&	15.704	&	0.250	&	0.311\\
\enddata
\end{deluxetable}
\end{center}

\clearpage
\begin{center}

\begin{deluxetable}{c c c c c c c c c c c c}
\tabletypesize{\footnotesize}
\tablecolumns{12}
\tablewidth{0pt}
\tablecaption{EW lists\label{line_list}}
\tablehead{
			
			\colhead{}						&
			\colhead{}						&
			\colhead{}						&
			\colhead{}						&
			\multicolumn{2}{c}{\textbf{57}}		& 
			\multicolumn{2}{c}{\textbf{79}}		&
			\multicolumn{2}{c}{\textbf{117}}    	&
			\colhead{}						&
			\colhead{}						\\
			
			\colhead {Ion}					&
			\colhead{$\lambda$ (\AA)}			&
			\colhead{E.P. (eV)}				&
			\colhead{log(g\textit{f})}			&
			\colhead{EW}					&
			\colhead{log $\epsilon$}			&
			\colhead{EW}					&
			\colhead{log $\epsilon$}			&
			\colhead{EW}					&
			\colhead{log $\epsilon$}			&
			\colhead{Method}				&
			\colhead{Ref.}					}

\startdata

\textbf{O I}	&	6300.30	&	0.000	&	-9.820	&	\nodata	&	7.07	&	\nodata	&	7.04	&	\nodata	&	7.15	&	synth	&	1\\
	&	7771.94	&	9.150	&	0.370	&	\nodata	&	\nodata	&	\nodata	&	\nodata	&	\nodata	&	\nodata	&	EW	&	2\\
	&	7774.17	&	9.150	&	0.220	&	\nodata	&	\nodata	&	\nodata	&	\nodata	&	\nodata	&	\nodata	&	EW	&	2\\
	&	7775.39	&	9.150	&	0.000	&	\nodata	&	\nodata	&	\nodata	&	\nodata	&	\nodata	&	\nodata	&	EW	&	2\\ \\
\textbf{Na I}	&	5889.95	&	0.000	&	-0.190	&	\nodata	&	\nodata	&	221.6	&	3.97	&	\nodata	&	\nodata	&	EW	&	3\\
	&	5895.92	&	0.000	&	0.110	&	205.5	&	4.19	&	194.2	&	3.94	&	\nodata	&	\nodata	&	EW	&	3\\
	&	8183.26	&	2.100	&	0.240	&	67.1	&	4.21	&	55.5	&	4.01	&	75.1	&	4.06	&	EW	&	3\\
	&	8194.82	&	2.100	&	0.490	&	111.9	&	4.53	&	82.5	&	4.10	&	112.6	&	4.27	&	EW	&	3\\ \\
\textbf{Mg I}	&	3829.36	&	2.710	&	-0.227	&	\nodata	&	\nodata	&	232.6	&	5.47	&	\nodata	&	\nodata	&	EW	&	4\\
	&	3832.31	&	2.710	&	0.125	&	\nodata	&	\nodata	&	\nodata	&	\nodata	&	\nodata	&	\nodata	&	EW	&	4\\
	&	4571.10	&	0.000	&	-5.623	&	105.9	&	5.54	&	106.3	&	5.41	&	158.9	&	5.59	&	EW	&	4\\
	&	5172.70	&	2.710	&	-0.393	&	\nodata	&	\nodata	&	\nodata	&	\nodata	&	135.6	&	5.86	&	EW	&	4\\
	&	5183.62	&	2.720	&	-0.167	&	\nodata	&	\nodata	&	\nodata	&	\nodata	&	\nodata	&	\nodata	&	EW	&	4\\
	&	5528.42	&	4.350	&	-0.498	&	100.3	&	5.54	&	117.1	&	5.74	&	\nodata	&	\nodata	&	EW	&	4\\ \\
\textbf{Al I}	&	3944.01	&	0.010	&	-0.638	&	\nodata	&	\nodata	&	215.2	&	4.16	&	\nodata	&	\nodata	&	EW	&	3\\
	&	3961.53	&	0.010	&	-0.336	&	\nodata	&	\nodata	&	177.2	&	3.43	&	\nodata	&	\nodata	&	EW	&	3\\ \\
\textbf{Si I}	&	3905.53	&	1.910	&	-1.041	&	293.6	&	5.55	&	255.2	&	5.36	&	\nodata	&	\nodata	&	EW	&	5\\ \\
\textbf{Ca I}	&	4226.74	&	0.000	&	0.244	&	\nodata	&	\nodata	&	\nodata	&	\nodata	&	\nodata	&	\nodata	&	EW	&	6\\
	&	5588.76	&	2.530	&	0.210	&	69.1	&	4.12	&	73.7	&	4.13	&	95.1	&	4.26	&	EW	&	6\\
	&	5857.46	&	2.930	&	0.230	&	51.1	&	4.30	&	43.0	&	4.14	&	66.6	&	4.31	&	EW	&	6\\

\tablecomments{The machine-readable version of the entire table is available in the online 
journal. A portion of the table is shown here for guidance concerning form and content. The abundances listed
in this table are given for corresponding individual transitions.}

\enddata

\end{deluxetable}
\end{center}

\clearpage
\begin{center}

\begin{deluxetable}{c c c c c c c c c c}
\tabletypesize{\footnotesize}
\tablecolumns{10}
\tablewidth{0pt}
\tablecaption{Atmospheric parameters\label{atm_params}}
\tablehead{
			\colhead{}							&
			\colhead{}							&
			\multicolumn{2}{c}{\textbf{PHOT}}		&
			\multicolumn{2}{c}{\textbf{SPEC}}		&
			\multicolumn{4}{c}{\textbf{COMB}}		\\
			
			\colhead{Star}				&
			\colhead{\textit{S/N}}			&
			\colhead{\teff}			         &
			\colhead{\logg}				&
			\colhead{\teff}   				&
			\colhead{\logg}				&
			\colhead{\teff}				&
			\colhead{\logg}                           &
			\colhead{[Fe/H]}			&
			\colhead{\vmicro}			}			

\startdata
\multicolumn{10}{c}{RGB parameters} \\																			
\hline																			
\textbf{57}	&	140	&	4626	&	1.54	&	4468	&	0.83	&	4626	&	1.25	&	-2.51	&	2.31	\\
\textbf{79}	&	130	&	4553	&	1.35	&	4418	&	0.50	&	4553	&	0.91	&	-2.52	&	2.41	\\
\textbf{117}$^{\mathrm{\dagger}}$	&	200	&	4310	&	0.75	&	\nodata	&	\nodata	&	4310	&	0.75	&	-2.52	&	2.33	\\
\textbf{160}$^{\mathrm{\dagger}}$	&	230	&	4168	&	0.65	&	\nodata	&	\nodata	&	4168	&	0.65	&	-2.52	&	2.33	\\
\textbf{172}	&	100	&	4687	&	1.69	&	4515	&	0.82	&	4687	&	1.26	&	-2.45	&	2.30	\\
\textbf{226}	&	90	&	4844	&	2.18	&	4655	&	1.22	&	4844	&	1.71	&	-2.51	&	2.05	\\
\textbf{256}	&	90	&	4789	&	2.02	&	4532	&	1.00	&	4789	&	1.67	&	-2.58	&	1.97	\\
\textbf{440}	&	100	&	4756	&	1.86	&	4505	&	0.98	&	4756	&	1.69	&	-2.47	&	2.04	\\
\textbf{450}	&	150	&	4513	&	1.18	&	4342	&	0.02	&	4513	&	0.51	&	-2.65	&	3.08	\\
\textbf{472}	&	150	&	4414	&	1.02	&	4305	&	0.25	&	4414	&	0.69	&	-2.53	&	2.38	\\
\textbf{481}	&	180	&	4247	&	0.77	&	4257	&	0.23	&	4247	&	0.18	&	-2.47	&	2.42	\\ \\
\hline																			
\multicolumn{10}{c}{RHB parameters} \\																			
\hline																			
\textbf{36}	&	90	&	5905	&	2.62	&	5532	&	1.72	&	5905	&	2.42	&	-2.46	&	3.01	\\
\textbf{47}	&	100	&	5724	&	2.55	&	5392	&	1.70	&	5724	&	2.35	&	-2.47	&	2.57	\\
\textbf{202}	&	90	&	6257	&	2.73	&	6180	&	2.30	&	6257	&	2.46	&	-2.31	&	3.45	\\
\textbf{334}	&	30	&	5756	&	2.57	&	5220	&	1.37	&	5756	&	2.40	&	-2.42	&	2.93	\\
\textbf{403}	&	100	&	5649	&	2.49	&	5198	&	1.30	&	5649	&	2.20	&	-2.49	&	2.71	\\
\textbf{454}	&	70	&	5972	&	2.59	&	5860	&	2.27	&	5972	&	2.53	&	-2.36	&	3.24	\\
\textbf{458}	&	80	&	6001	&	2.67	&	5670	&	2.71	&	6001	&	2.30	&	-2.48	&	3.36	\\
\textbf{533}	&	60	&	5792	&	2.58	&	5430	&	1.63	&	5792	&	2.33	&	-2.40	&	2.61	\\
\textbf{547}	&	100	&	6197	&	2.72	&	5885	&	2.03	&	6197	&	2.58	&	-2.37	&	3.81	\\ \\
\hline																			
\multicolumn{10}{c}{BHB parameters} \\																			
\hline																			
\textbf{170}	&	80	&	7848	&	3.24	&	7810	&	3.43	&	7848	&	3.50	&	-2.06	&	2.30	\\
\textbf{289}	&	80	&	7410	&	3.10	&	7680	&	3.05	&	7410	&	2.59	&	-2.58	&	1.97	\\
\textbf{324}$^{\mathrm{\dagger}}$	&	50	&	7792	&	3.25	&	\nodata	&	\nodata	&	7792	&	3.18	&	-2.25	&	2.17	\\
\textbf{337}	&	80	&	7700	&	3.20	&	8010	&	3.68	&	7700	&	3.24	&	-2.08	&	2.02	\\
\textbf{391}	&	100	&	7444	&	3.15	&	7497	&	3.11	&	7444	&	3.04	&	-2.29	&	2.39	\\

\enddata
\tablenotetext{\dagger}{These stars are represented with green markers in Figs. \ref{cmd_m68}, \ref{isochrone}, and \ref{isochrone_color}. See \ref{finalparams} for details.}
\end{deluxetable}
\end{center}

\clearpage
\begin{center}

\begin{deluxetable}{c c c c}
\tabletypesize{\footnotesize}
\tablecolumns{4}
\tablewidth{0pt}
\tablecaption{Parameter uncertainties\label{uncertainties}}
\tablehead{
			\colhead {Evolutionary stage}						&
			\colhead{ $\sigma_{\teff}$[K]}						&
			\colhead{ $\sigma_{\logg}\mathrm{[cm/s^{2}]}$}		&
			\colhead{ $\sigma_{\vmicro}$[km/s]}					}

\startdata
early RGB & 100 & 0.30 & 0.20 \\
late RGB & 75 & 0.34 & 0.25 \\
RHB & 150 & 0.30 & 0.40 \\
BHB & 200 & 0.35 & 0.20 \\

\enddata

\end{deluxetable}
\end{center}

\clearpage
\begin{center}

\begin{deluxetable}{c c c}
\tabletypesize{\footnotesize}
\tablecolumns{3}
\tablewidth{0pt}
\tablecaption{[Al/Fe] and [Na/Fe] NLTE abundances\label{al_nlte}}
\tablehead{
			\colhead{Star}						&
			\colhead{$\mathrm{[Al/Fe]_{NLTE}}$}	&
			\colhead{$\mathrm{[Na/Fe]_{NLTE}}$}		}

\startdata

79	&	0.26	&	0.08 \\
172	&	-0.34	&	 -0.26 \\
226	&	0.73 	&	0.63 \\
256	&	0.50 	&	0.30 \\
440	&	0.42 	&	-0.14 \\

\enddata

\end{deluxetable}
\end{center}

\clearpage
\begin{center}

\begin{deluxetable}{c c}
\tabletypesize{\footnotesize}
\tablecolumns{2}
\tablewidth{0pt}
\tablecaption{Comparison study references\label{abund_sources}}
\tablehead{
			\colhead{Element}				&
			\colhead{Reference}						}

\startdata

O I	&	\cite{carretta_na-o_2009} \\
Na I	&	\cite{carretta_na-o_2009} \\
Mg I	&	\cite{carretta_na-o_2009} \\
Al I	&	\nodata \\
Si I	&	\cite{carretta_na-o_2009} \\
Ca I	&	\cite{lee_chemical_2005} \& \cite{langer_spectroscopic_1998} \\
Sc II	&	\cite{lee_chemical_2005} \& \cite{langer_spectroscopic_1998} \\
Ti I	&	\cite{lee_chemical_2005} \& \cite{langer_spectroscopic_1998} \\
Ti II	&	\cite{lee_chemical_2005} \& \cite{langer_spectroscopic_1998} \\
Cr I	&	\cite{langer_spectroscopic_1998} \\
Cr II	&	\nodata \\
Mn I	&	\cite{lee_chemical_2005} \\
Mn II	&	\nodata \\
Fe I	&	\nodata \\
Fe II	&	\nodata \\
Co I	&	\cite{langer_spectroscopic_1998} \\
Cu I	&	\cite{lee_chemical_2005} \\
Zn I	&	\nodata \\
Sr II	&	\cite{langer_spectroscopic_1998} \\
Y II	&	\nodata \\
Zr II	&	\nodata \\
Ba II	&	\cite{lee_chemical_2005} \& \cite{langer_spectroscopic_1998} \\
La II	&	\cite{lee_chemical_2005} \\
Nd II	&	\nodata \\
Eu II	&	\cite{lee_chemical_2005} \\
Dy II	&	\nodata \\
Er II	&	\nodata \\
Yb II	&	\nodata \\

\enddata

\end{deluxetable}
\end{center}

\clearpage
\begin{center}

\begin{deluxetable}{c c c c c c c c}
\tabletypesize{\footnotesize}
\tablecolumns{8}
\tablewidth{0pt}
\tablecaption{Average abundances for individual stars\label{individual_abunds}}
\tablehead{
			\colhead {}				&
			\colhead{\textbf{57}}			& 
			\colhead{\textbf{79}}			&
			\colhead{\textbf{117}}		&
			\colhead{\textbf{160}}		&
			\colhead{\textbf{172}}		&
			\colhead{\textbf{226}}		&
			\colhead{\textbf{256}}		\\
			
			\colhead {Ion}						&
			\colhead{$\langle$[X/Fe]$\rangle$}		& 
			\colhead{$\langle$[X/Fe]$\rangle$}		& 
			\colhead{$\langle$[X/Fe]$\rangle$}		& 
			\colhead{$\langle$[X/Fe]$\rangle$}		& 
			\colhead{$\langle$[X/Fe]$\rangle$}		& 
			\colhead{$\langle$[X/Fe]$\rangle$}		& 
			\colhead{$\langle$[X/Fe]$\rangle$}		}

\startdata	
\textbf{Fe I}$^{\mathrm{\dagger}}$	&	-2.51	&	-2.52 	&	-2.39	        &	-2.49	        &	-2.45 	&	-2.51 	&	-2.58 	\\
\textbf{Fe II}$^{\mathrm{\dagger}}$	&	-2.51	&	-2.52 	&	-2.16 	&	-2.18 	&	-2.45 	&	-2.51 	&	-2.58	         \\		
\textbf{\ion{O}{1}}	&	0.80	&	\nodata	&	0.89	&	0.76	&	\nodata	&	\nodata	&	\nodata	\\
\textbf{\ion{Na}{1}}	&	0.31	&	0.08	&	0.27	&	0.10	&	-0.26	&	0.63	&	0.30	\\
\textbf{\ion{Mg}{1}}	&	0.45	&	0.46	&	0.64	&	0.51	&	0.33	&	0.00	&	0.34	\\
\textbf{\ion{Al}{1}}	&	\nodata	&	-0.14	&	\nodata	&	\nodata	&	-0.74	&	0.18	&	-0.05	\\
\textbf{\ion{Si}{1}}	&	0.55	&	0.37	&	\nodata	&	\nodata	&	0.53	&	0.47	&	0.53	\\
\textbf{\ion{Ca}{1}}	&	0.34	&	0.33	&	0.43	&	0.33	&	0.41	&	0.36	&	0.39	\\
\textbf{\ion{Ti}{1}}	&	0.09	&	0.15	&	0.08	&	-0.08	&	0.14	&	0.09	&	0.16	\\
\textbf{\ion{Ti}{2}}	&	0.33	&	0.28	&	0.72	&	0.75	&	0.25	&	0.30	&	0.41	\\
\textbf{\ion{Sc}{2}}	&	0.16	&	0.12	&	0.43	&	0.42	&	0.18	&	0.10	&	0.20	\\
\textbf{\ion{Cr}{1}}	&	-0.15	&	-0.08	&	\nodata	&	-0.06	&	-0.02	&	-0.17	&	0.01	\\
\textbf{\ion{Cr}{2}}	&	0.36	&	0.26	&	0.56	&	0.54	&	0.29	&	0.41	&	0.34	\\
\textbf{\ion{Mn}{1}}	&	-0.39	&	-0.37	&	-0.25	&	-0.26	&	-0.36	&	-0.36	&	-0.38	\\
\textbf{\ion{Mn}{2}}	&	\nodata	&	\nodata	&	\nodata	&	\nodata	&	0.08	&	\nodata	&	\nodata	\\
\textbf{\ion{Co}{1}}	&	\nodata	&	-0.19	&	-0.04	&	\nodata	&	0.28	&	0.22	&	0.37	\\
\textbf{\ion{Cu}{1}}	&	\nodata	&	\nodata	&	-0.61	&	-0.28	&	\nodata	&	\nodata	&	\nodata	\\
\textbf{\ion{Zn}{1}}	&	0.23	&	0.12	&	0.21	&	0.29	&	0.09	&	0.25	&	0.22	\\
\textbf{\ion{Sr}{2}}	&	-0.45	&	-0.78	&	-0.35	&	-0.19	&	-0.85	&	-0.57	&	-0.41	\\
\textbf{\ion{Y}{2}}	&	-0.30	&	-0.45	&	-0.36	&	-0.32	&	-0.48	&	-0.41	&	-0.30	\\
\textbf{\ion{Zr}{2}}	&	-0.11	&	-0.13	&	-0.04	&	0.04	&	-0.26	&	-0.15	&	0.01	\\
\textbf{\ion{Ba}{2}}	&	-0.26	&	-0.31	&	-0.18	&	-0.10	&	-0.44	&	-0.38	&	-0.22	\\
\textbf{\ion{La}{2}}	&	-0.06	&	-0.18	&	-0.16	&	-0.15	&	-0.29	&	0.13	&	-0.05	\\
\textbf{\ion{Nd}{2}}	&	0.20	&	0.07	&	0.07	&	0.31	&	0.05	&	0.11	&	0.23	\\
\textbf{\ion{Eu}{2}}	&	0.26	&	0.13	&	\nodata	&	0.32	&	0.07	&	0.15	&	0.47	\\
\textbf{\ion{Dy}{2}}	&	0.45	&	0.40	&	0.43	&	0.69	&	0.31	&	0.40	&	0.33	\\
\textbf{\ion{Er}{2}}	&	0.22	&	0.08	&	\nodata	&	0.46	&	0.31	&	0.22	&	0.43	\\
\textbf{\ion{Yb}{2}}	&	0.22	&	0.12	&	\nodata	&	0.11	&	0.07	&	\nodata	&	0.15	\\

\enddata

\tablenotetext{\dagger}{These abundances are [X/H], instead of [X/Fe].}

\tablecomments{The machine-readable version of the entire table is available in the online 
journal. A portion of the table is shown here for guidance concerning form and content.}

\end{deluxetable}
\end{center}

\clearpage
\begin{center}

\begin{deluxetable}{c c c c c c}
\tabletypesize{\footnotesize}
\tablecolumns{6}
\tablewidth{0pt}
\tablecaption{Average abundances for the individual evolutionary stages and the complete cluster M68\label{evolution_abund}}
\tablehead{
			\colhead {}							&
			\colhead{\textbf{RGB}}					& 
			\colhead{\textbf{RHB}}					&
			\colhead{\textbf{BHB}}					&
			\multicolumn{2}{c}{\textbf{M68}}			\\
			
			\colhead{Ion}				&
			\colhead{$\langle$[X/Fe]$\rangle$}		& 
			\colhead{$\langle$[X/Fe]$\rangle$}		& 
			\colhead{$\langle$[X/Fe]$\rangle$}		& 
			\colhead{$\langle$[X/Fe]$\rangle$}		& 
			\colhead{$\sigma_{\langle\mathrm{[X/Fe]}\rangle}$}	}
			
\startdata
\textbf{Fe I}$^{\mathrm{\dagger}}$	&	-2.51 	&	-2.42 	&	-2.25 	&	-2.423	&	0.141	\\
\textbf{Fe II}$^{\mathrm{\dagger}}$	&	-2.45	&	-2.42	&	-2.25	&	-2.402	&	0.156	\\
\textbf{\ion{O}{1}}	&	0.69	&	0.52	&	0.31	&	0.506	&	0.275	\\
\textbf{\ion{Na}{1}}	&	0.20	&	0.06	&	0.35	&	0.165	&	0.264	\\
\textbf{\ion{Mg}{1}}	&	0.36	&	0.35	&	0.35	&	0.353	&	0.170	\\
\textbf{\ion{Al}{1}}	&	-0.23	&	-0.40	&	-0.07	&	-0.276	&	0.369	\\
\textbf{\ion{Si}{1}}	&	0.40	&	0.16	&	-0.13	&	0.207	&	0.238	\\
\textbf{\ion{Ca}{1}}	&	0.36	&	0.43	&	0.24	&	0.360	&	0.111	\\
\textbf{\ion{Ti}{1}}	&	0.08	&	0.43	&	\nodata	&	0.215	&	0.191	\\
\textbf{\ion{Ti}{2}}	&	0.42	&	0.43	&	0.49	&	0.438	&	0.134	\\
\textbf{\ion{Sc}{2}}	&	0.21	&	0.37	&	0.22	&	0.270	&	0.134	\\
\textbf{\ion{Cr}{1}}	&	-0.08	&	0.19	&	0.18	&	0.028	&	0.165	\\
\textbf{\ion{Cr}{2}}	&	0.36	&	0.44	&	0.33	&	0.384	&	0.111	\\
\textbf{\ion{Mn}{1}}	&	-0.35	&	-0.40	&	\nodata	&	-0.366	&	0.150	\\
\textbf{\ion{Mn}{2}}	&	-0.11	&	0.05	&	\nodata	&	0.015	&	0.340	\\
\textbf{\ion{Fe}{1}}	&	0.01	&	0.00	&	0.10	&	0.027	&	0.105	\\
\textbf{\ion{Fe}{2}}	&	0.06	&	0.00	&	0.10	&	0.049	&	0.137	\\
\textbf{\ion{Co}{1}}	&	0.07	&	0.25	&	\nodata	&	0.146	&	0.188	\\
\textbf{\ion{Cu}{1}}	&	-0.54	&	\nodata	&	\nodata	&	-0.540	&	0.181	\\
\textbf{\ion{Zn}{1}}	&	0.19	&	0.45	&	\nodata	&	0.226	&	0.133	\\									
\textbf{\ion{Sr}{2}}	&	-0.56	&	-0.16	&	-0.78	&	-0.444	&	0.296	\\
\textbf{\ion{Y}{2}}	&	-0.42	&	-0.29	&	\nodata	&	-0.363	&	0.110	\\
\textbf{\ion{Zr}{2}}	&	-0.10	&	0.19	&	\nodata	&	0.032	&	0.198	\\
\textbf{\ion{Ba}{2}}	&	-0.28	&	-0.17	&	-0.55	&	-0.287	&	0.163	\\
\textbf{\ion{La}{2}}	&	-0.13	&	0.23	&	\nodata	&	-0.098	&	0.164	\\
\textbf{\ion{Nd}{2}}	&	0.13	&	0.37	&	\nodata	&	0.166	&	0.132	\\
\textbf{\ion{Eu}{2}}	&	0.19	&	0.29	&	\nodata	&	0.227	&	0.141	\\
\textbf{\ion{Dy}{2}}	&	0.38	&	0.65	&	\nodata	&	0.485	&	0.215	\\
\textbf{\ion{Er}{2}}	&	0.35	&	0.77	&	\nodata	&	0.523	&	0.297	\\
\textbf{\ion{Yb}{2}}	&	0.15	&	0.41	&	\nodata	&	0.296	&	0.173	\\
\enddata

\tablenotetext{\dagger}{These abundances are [X/H], instead of [X/Fe].}

\end{deluxetable}
\end{center}

\clearpage
\begin{center}

\begin{deluxetable}{c c c c c}
\tabletypesize{\footnotesize}
\tablecolumns{5}
\tablewidth{0pt}
\tablecaption{Abundance uncertainties\label{abund_uncert}}
\tablehead{
			\colhead{}					&
			\colhead{472 (RGB tip)}		&
			\colhead{172 (RGB)}			&
			\colhead{36 (RHB)}			&
			\colhead{337 (BHB)}			\\

			\colhead{Species}		&
			\colhead{$\sigma$}		&
			\colhead{$\sigma$}		&
			\colhead{$\sigma$}		&
			\colhead{$\sigma$}		}

\startdata
\multicolumn{5}{c}{EW species} \\
\hline
\textbf{O I}	&	$\pm$	0.08	&	$\pm$	0.08	&	$\pm$	0.14	&	$\pm$	0.19	\\
\textbf{Na I}	&	$\pm$	0.18	&	$\pm$	0.13	&	$\pm$	0.20	&	$\pm$	0.11	\\
\textbf{Mg I}	&	$\pm$	0.27	&	$\pm$	0.17	&	$\pm$	0.09	&	$\pm$	0.29	\\
\textbf{Al I}	&	$\pm$	0.11	&	$\pm$	0.18	&	$\pm$	0.09	&	$\pm$	0.13	\\
\textbf{Si I}	&	$\pm$	0.15	&	$\pm$	0.21	&	$\pm$	0.15	&	$\pm$	0.09	\\
\textbf{Ca I}	&	$\pm$	0.12	&	$\pm$	0.10	&	$\pm$	0.11	&	$\pm$	0.12	\\
\textbf{Sc II}	&	$\pm$	0.09	&	$\pm$	0.18	&	$\pm$	0.13	&	$\pm$	0.07	\\
\textbf{Ti I}	&	$\pm$	0.16	&	$\pm$	0.15	&	$\pm$	0.10	&		\nodata	\\
\textbf{Ti II}	&	$\pm$	0.08	&	$\pm$	0.08	&	$\pm$	0.11	&	$\pm$	0.11	\\
\textbf{Cr I}	&	$\pm$	0.10	&	$\pm$	0.08	&	$\pm$	0.20	&	$\pm$	0.15	\\
\textbf{Cr II}	&	$\pm$	0.05	&	$\pm$	0.08	&	$\pm$	0.11	&	$\pm$	0.09	\\
\textbf{Mn I}	&	$\pm$	0.10	&	$\pm$	0.23	&	$\pm$	0.08	&		\nodata	\\
\textbf{Mn II}	&		\nodata	&	$\pm$	0.24	&	$\pm$	0.25	&		\nodata	\\
\textbf{Fe I}	&	$\pm$	0.12	&	$\pm$	0.13	&	$\pm$	0.13	&	$\pm$	0.18	\\
\textbf{Fe II}	&	$\pm$	0.16	&	$\pm$	0.14	&	$\pm$	0.10	&	$\pm$	0.13	\\
\textbf{Co I}	&	$\pm$	0.18	&	$\pm$	0.26	&	$\pm$	0.13	&		\nodata	\\
\textbf{Cu I}	&	$\pm$	0.08	&		\nodata	&		\nodata	&		\nodata	\\
\textbf{Zn I}	&	$\pm$	0.07	&	$\pm$	0.09	&	$\pm$	0.13	&		\nodata	\\ \\

\multicolumn{5}{c}{Synth species} \\
\hline
\textbf{Sr II}	&	$\pm$	0.15	&	$\pm$	0.14	&	$\pm$	0.13	&	$\pm$	0.13	\\
\textbf{Y II}	&	$\pm$	0.11	&	$\pm$	0.16	&	$\pm$	0.17	&		\nodata	\\
\textbf{Zr II}	&	$\pm$	0.09	&	$\pm$	0.12	&	$\pm$	0.09	&		\nodata	\\
\textbf{Ba II}	&	$\pm$	0.10	&	$\pm$	0.08	&	$\pm$	0.13	&	$\pm$	0.13	\\
\textbf{La II}	&	$\pm$	0.10	&	$\pm$	0.09	&	$\pm$	0.18	&		\nodata	\\
\textbf{Nd II}	&	$\pm$	0.16	&	$\pm$	0.12	&	$\pm$	0.17	&	$\pm$	0.16	\\
\textbf{Eu II}	&	$\pm$	0.09	&	$\pm$	0.09	&	$\pm$	0.20	&		\nodata	\\
\textbf{Dy II}	&	$\pm$	0.14	&	$\pm$	0.12	&	$\pm$	0.17	&		\nodata	\\
\textbf{Er II}	&	$\pm$	0.16	&	$\pm$	0.09	&	$\pm$	0.14	&		\nodata	\\
\textbf{Yb II}	&	$\pm$	0.14	&	$\pm$	0.10	&	$\pm$	0.10	&	$\pm$	0.17	\\

\enddata
\end{deluxetable}
\end{center}

\end{document}